# Non-Hermitian Topology and Exceptional-Point Geometries

Kun Ding[1], Chen Fang[2,3,4], Guancong Ma[5]

[1] Department of Physics, State Key Laboratory of Surface Physics, and Key Laboratory of Micro and Nano Photonic Structures (Ministry of Education), Fudan University, Shanghai 200438, China

[2] Beijing National Laboratory for Condensed Matter Physics, Institute of Physics, Chinese Academy of Sciences, Beijing 100190, China

[3] Songshan Lake Materials Laboratory, Dongguan, Guangdong 523808, China

[4] Kavli Institute for Theoretical Sciences, Chinese Academy of Sciences, Beijing 100190, China

[5] Department of Physics, Hong Kong Baptist University, Kowloon Tong, Hong Kong, China

Emails: kunding@fudan.edu.cn, cfang@iphy.ac.cn, phgcma@hkbu.edu.hk

**Abstract.** Non-Hermitian theory is a theoretical framework that excels at describing open systems. It offers a powerful tool in the characterization of both the intrinsic degrees of freedom (DOFs) of a system and the interactions with the external environment. The non-Hermitian framework consists of mathematical structures that are fundamentally different from those of Hermitian theories. These structures not only underpin novel approaches for precisely tailoring non-Hermitian systems for applications but also give rise to topologies not found in Hermitian systems. In this paper, we comprehensively review non-Hermitian topology by establishing its relationship with the behaviors of complex eigenvalues and biorthogonal eigenvectors. Special attentions are given to exceptional points – branch-point singularities on the complex eigenvalue manifolds that exhibit non-trivial topological properties. We also discuss recent developments in non-Hermitian band topology, such as the non-Hermitian skin effect and non-Hermitian topological classifications.

**Key points.**

- A non-Hermitian Hamiltonian that describes an open system generically has complex eigenvalues, which must be studied on the complex plane, which leads to the emergence of eigenvalue topology, or spectral topology. This additional "layer" of topology is a unique feature for non-Hermitian systems.
- Spectral topology fundamentally affects the parallel-transport behaviors of eigenvectors of a non-Hermitian Hamiltonian.
- Exceptional points are branch singularities on non-Hermitian eigenvalue manifolds, and exhibit exotic topological phenomena associated with the winding of eigenvalues and eigenvectors.

- The confluence of non-Hermiticity and band topology generates new phenomena such as the non-Hermitian skin effect, which is characterized by non-Bloch band theory and the re-establishment of bulk-boundary correspondence. It also ramifies the possible symmetry classes, thereby expanding the classifications of topological bands.

**Introduction.** Quantum mechanics is built on Hermitian operators satisfying $H = H^\dagger$ – a condition that guarantees the existence of real eigenvalues and an orthonormal set of eigenvectors. This mathematical structure is ideal for analyzing the steady states in closed systems, in which both energy and probability are conserved. However, the necessary mathematical condition for real eigenvalues is that an operator obeys pseudo-Hermiticity[1,2], i.e., $\eta H \eta^{-1} = H^\dagger$, where $\eta$ is a Hermitian invertible operator, instead of the more stringent Hermitian condition. In the late 1990s, parity-time symmetry is identified as a form of pseudo-Hermiticity with a Hamiltonian that satisfies $(\mathcal{PT})H(\mathcal{PT})^{-1} = H^\dagger$, where $\mathcal{P}$ and $\mathcal{T}$ are respectively the parity and time-reversal operators[3–9]. Such systems soon became a physical reality, enabled by the precise engineering of decay rate in waves, optics, and QED cavities[10–16]. Although the eigenvalues remain real, these systems are no longer closed as they exchange energy in a balanced manner with the external environment. Once this balance is broken, e.g., when the gain or loss is too high, complex eigenvalues appear. This change is marked by a spectral singularity, denoted an exceptional point (EP)[17–19]. Thanks to the developments in metamaterials, photonics, acoustics, etc., this area rapidly became a new frontier of physics research, and a wide diversity of applications have emerged. These developments have been comprehensively reviewed in refs. [20–27].

The implications of complex eigenvalues go far beyond exotic wave phenomena. Because complex functions generically map to a two-dimensional (2D) manifold, it is possible for the complex-eigenvalue manifold, denoted $\mathcal{M}_E(z)$ with $z \in \mathbb{C}$ as system parameter(s), to be topologically different from its underlying manifold, that is, the parameter space. It is well recognized that topology plays a central role in determining the general behaviors of many physical systems – as exemplified by the vivid development of topological insulators. Yet all profound conclusions, ranging from the bulk-edge correspondence to the topological classifications, are based on Hermitian formalism, wherein "topology" studies the behaviors of eigenvectors, or wavefunctions, in the parameter space. The new possibilities lying within the non-Hermitian topology, especially those associated with the complex eigenvalues, which we denoted as spectral topology, have thus been the subject of intense research in the past few years.

This review article concisely anatomizes and surveys the different aspects of non-Hermitian topology and provides assessable examples to guide those who hope to enter the field. More experienced readers may also find ref.[26] useful, which offer a more mathematics-focused account of the topic. Ref. [28] is also an excellent review with an emphasis on non-Hermitian topology in periodic systems. This article is organized as follows. We begin by describing the two fundamental aspects of a Hamiltonian—its eigenvalues and eigenvectors. We then detail the complex eigenvalues of a non-Hermitian Hamiltonian and the concept of spectral areas or point gaps, which leads to the emergence of spectral topology and the definition of eigenvalue winding numbers (EWNs). An EP is subsequently defined as a branch point on $\mathcal{M}_E(z)$ and its topological characteristics are analyzed using the EWNs. The properties of non-Hermitian eigenvectors are then discussed, with a focus on the biorthogonality and defectiveness in the vicinity of an EP, followed by an examination of the parallel transport of non-Hermitian eigenvectors around an EP and a description of the biorthogonal Berry phase as an eigenvector winding number (VWN). We then present geometries formed by the EPs followed by higher-order EPs, and introduce important results such as non-Abelian state permutations. Finally, we examine non-Hermitian band topology, and detail the physics and topological origin of the non-Hermitian skin effect (NHSE) and the classification of non-Hermitian topological bands.

**Complex eigenvalues and spectral topology.** We begin by considering a Hermitian Hamiltonian $H(z)$, where $z$ denotes system parameter(s). We first set $z \in \mathbb{R}^d$ with $d$ being the dimensionality of the parameter space. (This condition is for the ease of discussion, and note that in general $H(z)$ can be Hermitian for complex $z$.) Hermiticity dictates that each eigenvalue $E(z) \in \mathbb{R}$ is confined to the real axis, so the eigenvalue manifold is also $d$-dimensional. When $z$ is varied according to some path $C_z$, its image on $E(z)$, denoted $C_E^H$, is always a homeomorphism of $C_z$, as shown in Fig. 1a. For example, if $z$ is the Bloch wavevector $k$ of a one-dimensional (1D) periodic system and the path $C_k$ traverses the Brillouin zone (BZ), then both $C_k$ and $C_E^H$ are homeomorphisms of an $S^1$ (a circle). That is, the Hermitian eigenvalue manifold is a homeomorphism of the system's base manifold. Therefore, interesting topology only emerges in Hermitian systems when the evolution of eigenvectors (or equivalently speaking, wavefunctions) —which are fiber bundles adhered to $C_E^H$—is further analyzed. In physical terms, the holonomy of $z$ induces a gauge-invariant Berry phase that can be quantized under certain conditions. A quantized Berry phase then constitutes the foundation of Hermitian topology[29].

Next, we examine a non-Hermitian Hamiltonian $H(z)$, where the system parameter(s) is, in general, $z \in \mathbb{C}$. Note that $H(z)$ can be non-Hermitian even if $z$ is entirely real, and this does not affect the following discussions. The eigenvalues are $E(z) \in \mathbb{C}$, which expands the dimensionality of the $\mathcal{M}_E(z)$. This simple difference between Hermitian and non-Hermitian realms means that the variation of $z$ can lead to a rich diversity of $C_E$ on the complex energy plane (Fig. 1a). In the Hermitian example, a closed path $C_z$, suppressed to a segment in Fig. 1a, maps to the curves $C_E^H$ in the spectrum, whereas in the non-Hermitian case, $C_z$ generally maps to a loop $C_E$ enclosing a spectral area on the complex energy plane. In other words, $C_E$ separate the complex-energy plane into disjoint areas. It follows that an EWN, $\mathcal{W}_E$, can naturally be defined

$$\mathcal{W}_E = \frac{1}{2\pi i} \oint_{C_z} d\vec{z} \cdot \nabla_z \ln \det[H(z) - E_r I], \quad (1)$$

where $E_r$ is an arbitrary reference energy, and $I$ is an identity matrix with the same dimension as $H(z)$. Equation (1) is the closed-loop integral of a gradient field, and according to Stokes' theorem this integral must vanish unless $C_z$ encloses singularities. This can be seen by considering a one-state non-Hermitian Hamiltonian $H_1(z)$, such that Eq. (1) becomes

$$\mathcal{W}_E = \frac{1}{2\pi i} \oint_{C_z} \frac{\partial E(z)/\partial z}{E(z) - E_r} dz. \quad (2)$$

According to Cauchy's argument principle, $\mathcal{W}_E = Z - P$, where $Z$ and $P$ are the number of zeros and poles of $E(z) - E_r$ inside $C_E$, respectively. Thus, the simplest condition for $\mathcal{W}_E$ to be non-zero is for $C_E$ to enclose a nonvanishing spectral area, such that an interior point can be selected as $E_r$. In this case, $E(z)$ encloses a spectral area, or equivalently, forms a "point gap."[30,31] Apparently, such an EWN is always trivial for Hermitian eigenvalues. However, in some non-Hermitian cases, such as when the pseudo-Hermitian condition holds, a complex $C_E$ can still enclose no spectral area such that $\mathcal{W}_E = 0$. Two bands $E_1(z)$ and $E_2(z)$ are regarded as isolated, or "line-gapped,"[30–32] when a line can be drawn on the complex energy plane to separate $C_{E_1}$ and $C_{E_2}$, as shown in Fig. 1a. Generically, the line can always be made perpendicular to the real or the imaginary axis via a rotation. Moreover, when the line is perpendicular to the real (imaginary) axis, the non-Hermitian bands separated by the line can be "flattened" to points on the real (imaginary) axis by a similarity transformation that connects the non-Hermitian system to a Hermitian (anti-Hermitian) system. More discussions on this idea are in the section "Classification of non-Hermitian topological bands."

Although the extension of eigenvalues to the complex energy plane may seem straightforward, it necessitates a re-examination of previous conclusions drawn from Hermitian theories. For instance, the generalization of degeneracy to non-Hermitian situations requires both the real and imaginary parts of energies to be identical. This condition occurs at an EP, which requires the existence of at least two bands. We therefore consider the following two-state Hamiltonian

$$H_2(z) = \begin{pmatrix} 0 & t \\ t & z \end{pmatrix}. \tag{3}$$

When $t, z \in \mathbb{R}$, Eq. (3) is Hermitian. It becomes non-Hermitian when either parameter is subjected to analytic continuation. Here, we let $z = x + iy$ with $x, y \in \mathbb{R}$ and $i^2 = -1$, and we set $t = 1$. There are a multitude of ways to realize such parameters in different systems. For example, $x$ may be realized as detuned onsite energy, and $iy$ as onsite loss (gain). The eigenvalues are

$$E_{+,-}(z) = \frac{1}{2}\left(z \pm \sqrt{4 + z^2}\right). \tag{4}$$

Equation (4) contains a square root of a complex function, which is a one-to-two map (Fig. 1b). Bernhard Riemann described the images of a complex square root by splitting the complex plane into two sheets that connect at a branch cut (Fig. 1c). In the context of non-Hermitian physics, the endpoint of a branch cut is a branch point that is precisely an EP. The two EPs of Eq. (4) are located at $z = \pm 2i$, denoted by the red dots in Fig. 1d. Figures 1e and 1f show the real and imaginary parts, respectively, of $E_{+,-}(z)$ near the EP at $z = 2i$. The eigenvalues are identical at the EP, that is $E_+ = E_-$. Expand the parameter $z$ near the EP as $z \cong \pm 2i\left(1 + \frac{1}{2}\delta z\right)$ with $\delta z$ being small, and the eigenvalue splitting follows $\Delta E = |E_+ - E_-| \cong 2|\sqrt{\delta z}|$ up to the leading order. This is the generic property of an EP formed by two coalescing states, hence the name a square-root EP or an order-2 EP. Such a square-root dependence essentially amplifies any small parametric variations – a useful property for sensor applications, wherein a resonant system operates in close vicinity of an EP, such that the eigenfrequencies of the relevant non-Hermitian modes become highly sensitive to perturbations to the system[33–36]. An alternative way to identify an EP is by the vanishing of the discriminant of the characteristic polynomial of Eq. (3), denoted $\Delta$, which is generally complex. Hence the vanishing condition gives two equations, and their common roots are the EPs, as shown in Fig. 1d.

As mentioned, the two eigenvectors at an EP are identical. This means that, in Eq. (3), $|\psi_{EP}\rangle = (\mp i, 1)^T$ for the EP at $= \pm 2i$, and the Hamiltonian becomes a defective matrix. This form of $|\psi_{EP}\rangle$ is similar to that of the Jones vector describing circular polarization of light, and is used to assign chirality to the EP[37–39]. This property fundamentally distinguishes an EP from a Hermitian degenerate point (or, a diabolic point), as the eigenvectors remain complete and orthogonal at the latter.

In Figs. 1e and 1f, it is clear that the geometry of the eigenvalue surface is rather unusual – they do not resemble any 2D surfaces that can be embedded in a 3D Euclidean space. Perhaps the most striking feature is that the two eigenvalues are smoothly connected when the system is continuously driven across a branch cut. Strictly, this means that one can no longer label the states with their eigenvalues. However, for the convenience of discussion and when no ambiguity is present, it is acceptable to count the states using the real part of the eigenvalues. Figures 1g–i show the $C_E$ of the two eigenvalues as indicated in the corresponding inset, which plots the eigenvalues' real parts. In Fig. 1g, in which $C_E$ crosses neither the EP nor the branch cut, the two bands form two spectral areas separated by a line gap, and the eigenvalues form braids that map to an unlink (or two unknots). In Fig. 1h, in which the loop crosses the EP, the line gap closes and the two spectral areas touch at the EP (Fig. 1h). Further shifting $C_E$ such that it encloses the EP and crosses the branch cut, the two eigenvalues together form a single point gap, and the eigenvalue braids form one unknot (Fig. 1i). This is a "gap-closing" maneuver unique to non-Hermitian bands: the two spectral areas merge into a single one. The EWN defined in Eq. (1) is suitable to describe this process. When the EP's energy is chosen as the reference $E_r$, the EWN is equivalent to

$$\mathcal{W}_E = v_\pm + v_\mp, \tag{5}$$

where $v_{\pm(\mp)} = -\frac{1}{2\pi}\oint_{C_z} d\vec{z} \cdot \nabla_z [\arg(E_{+(-)} - E_{-(+)})]$ is called eigenvalue vorticity[32,40,41]. In some studies, Eq. (5) is also denoted the discriminant number and it can be generalized to multi-band systems[41]. The merging of spectral areas shown in Figs. 1g–i corresponds to a jump of $\mathcal{W}_E$ from 0 to –1, indicating a topological transition in terms of the spectral topology. Note that the EWN here describes the combined winding of both eigenvalues. Alternatively, the EWN characterizes the winding of one eigenvalue on $\mathcal{M}_E(z)$ until $C_E$ forms a closed loop, which requires the corresponding $C_z$ to complete two cycles of winding around $z = 2i$. Such a difference between $C_E$ and $C_z$ is the consequence of the presence of EP and the corresponding branch cut, and it indicates that $\mathcal{M}_E(z)$ is topologically different from the base manifold. Indeed, the very existence of an EP as a singularity implies a nontrivial fundamental group for $\mathcal{M}_E(z)$, that is, loops enclosing an EP cannot smoothly contract to a point, whereas loops that do not enclose an EP are equivalent to a point.

Eigenvalue winding behaviors are observable by experimentally following the complex $E$ as a loop is traced out by system parameter(s). Notable examples are the winding of a single band based on a one-band non-Hermitian model in optics with synthetic periodicity[42], and the simultaneous winding of two bands around EPs in acoustics[43]. Recent studies have shown that braids or knot theories are an alternative way to characterize the winding of multiple complex eigenvalues[44–46]. This can be seen by extending $C_E$ along the argument of $z$, which enables the windings of $E_\pm(z)$ to map to an unlink (Fig. 1g), a chain (Fig. 1h), and an unknot (Fig. 1i). More sophisticated knot and link structures, such as a Hopf link and a trefoil knot, can be realized by designing the encircling paths and (or) the geometry of $\mathcal{M}_E(z)$ [45–47].

**Non-Hermitian eigenvectors and their topology.** The eigenvectors of non-Hermitian Hamiltonians and their evolutionary behaviors are also drastically different from those of their Hermitian counterparts. First, the left and right eigenvectors of a non-Hermitian Hamiltonian, defined as $H|\psi^R\rangle = E|\psi^R\rangle$ and $\langle\psi^L|H = E\langle\psi^L|$, respectively, are not the Hermitian conjugate of each other, namely, $|\psi^R\rangle^\dagger \neq \langle\psi^L|$. This means that the left and right eigenvectors must be obtained separately. In practice, it is convenient to compute the left eigenvectors by first obtaining the right eigenvectors of $H^\dagger (\neq H)$, i.e. $H^\dagger|\psi'^R\rangle = E^*|\psi'^R\rangle$, and then taking the Hermitian conjugate to obtain $\langle\psi^L| = |\psi'^R\rangle^\dagger$.

Non-Hermitian right eigenvectors of different states are skewed instead of orthogonal. Such skewness underlies phenomena such as NHSE, as will be discussed in the section "Non-Hermitian band topology and skin effects." Non-Hermitian eigenvectors are normalized using a bilinear product of the left and right eigenvectors[48,49],

$$|\bar\psi^R\rangle = \frac{|\psi^R\rangle}{\sqrt{\langle\psi^L|\psi^R\rangle}}, \qquad \langle\bar\psi^L| = \frac{\langle\psi^L|}{\sqrt{\langle\psi^L|\psi^R\rangle}}, \tag{8}$$

where the denominators are generally complex. The bilinear-normalized eigenvectors then satisfy the biorthonormal condition

$$\langle\bar\psi^L_m|\bar\psi^R_n\rangle = \delta_{mn}, \tag{9}$$

where $m, n$ label the states. Special care is required to handle the eigenvectors as the system approaches an EP, as they become heavily skewed and ultimately identical (Fig. 1j). For example, the bilinear-normalized right eigenvectors of Eq. (3) are

$$|\bar\psi^R_{+,-}\rangle = \frac{1}{\sqrt{4 + \left(z \pm \sqrt{4+z^2}\right)^2}} \begin{pmatrix} -z \mp \sqrt{4+z^2} \\ 2 \end{pmatrix}. \tag{10}$$

Expanding near the EP at $z = -2i$ as $z = -2i\left(1 + \frac{1}{2}\delta z\right)$, we obtain

$$|\bar\psi^R_{+,-}\rangle = \frac{1}{i\sqrt{2}\delta z^{1/4}}\left[\begin{pmatrix} i \\ 1 \end{pmatrix} + \begin{pmatrix} \mp i\sqrt{\delta z} \\ 0 \end{pmatrix}\right]. \tag{11}$$

When $\delta z \sim 0$, the second term vanishes and the two states become identical, leading to the defective character of the Hamiltonian at the EP. The defectiveness is also revealed by casting the Hamiltonian matrix at the EP into a Jordan canonical form, which has null eigenvectors[26,50]. Surprisingly, null eigenvectors at an EP can still be observed by specific excitation that suppresses the non-defective eigenvectors[51]. The defectiveness is responsible for a specific type of bound states, which can result in negative entanglement entropy[52,53].

Equation (11) also implies that the splitting of eigenvectors follows the factor $\sqrt{\delta z}$, which is quantified using phase rigidity[54,55], defined as

$$r = \frac{\langle \bar{\psi}^L | \bar{\psi}^R \rangle}{\langle \bar{\psi}^R | \bar{\psi}^R \rangle}, \tag{12}$$

which gives $r_{+,-} = \delta z^{1/2}$. The power of $\delta z$ is called the critical exponent $s$, indicating the splitting of eigenvectors also follows a square-root dependence similar to that of the eigenvalues. However, this is not a general rule; it has been shown that, even in order-2 EPs, the splitting of eigenvalues can follow different parametric dependences[32,56]. It is also possible that the splitting follows different dependences in the critical exponent $s$ when an EP is approached from different parametric directions, meaning that the EP appears to be "anisotropic."[56]

The critical exponent of phase rigidity, $s = 1/2$, can thus be treated as a fractional VWN[55,57]. VWNs are associated with the parallel transport of eigenvectors, which are fiber bundles adhered to the $C_E$, which is on $\mathcal{M}_E(z)$. The spectral topology therefore governs the evolution of the eigenvectors[58,59]. An immediate consequence is that eigenvectors can swap by evolving into one another when the trajectory crosses a branch cut. This behavior is the underlying mechanism of non-Hermitian state permutations and is captured by a multi-band Berry phase, which is essentially an $N \times N$ unitary matrix $\boldsymbol{U}(\boldsymbol{N})$ with elements

$$U_{mn} = \mathcal{P} \exp\left(i \oint_{C_z} A_{mn} dz\right), \tag{13}$$

where $A_{mn} = i\langle \bar{\psi}^L_m | \nabla_z | \bar{\psi}^R_n \rangle$ is the biorthogonal Berry connection, and $\mathcal{P}$ is the path-ordering operator. In practice, $\boldsymbol{U}(\boldsymbol{N})$ is often evaluated by segmenting $C_z$ into $L$ discretized steps

$$\boldsymbol{U}(\boldsymbol{N}) = \prod_{l=1}^{L-1} \boldsymbol{M}(z_l, z_{l+1}), \tag{14}$$

wherein $M_{mn} = \langle \bar{\psi}^L_m(z_l) | \bar{\psi}^R_n(z_{l+1}) \rangle$. We can further obtain a phase factor

$$\theta = -\text{Im}[\ln \det \boldsymbol{U}(\boldsymbol{N})]. \tag{15}$$

This phase factor is the gauge-invariant multi-band Berry phase[60]. An equivalent approach is to follow the evolution of a particular eigenvector along $C_E$ until the initial state is restored up to a phase factor that equals $\theta$. If a state exchange occurs, the eigenvector may recover after $n$ complete cycles. This implies that $\theta$ can also be computed using the single-band Wilson loop approach, that is

$$\theta = -\text{Im}\left\{\ln\left[\prod_{l=1}^{nL-1} \langle \bar{\psi}^L_m(z_l) | \bar{\psi}^R_m(z_{l+1}) \rangle\right]\right\}. \tag{16}$$

Figure 1k traces an eigenvector of Eq. (3), $|\bar{\psi}_+^R\rangle$, where the evolution path is identical to the $C_z$ shown in Fig. 1i, namely, it encloses the EP. The corresponding $C_E$ crosses the branch cut once per cycle hence it is closed after $n = 2$ completely cycles of $C_z$. In addition, parallel transport is numerically imposed at each step[60,61]. In Fig. 1k, we can see that it indeed takes two complete cycles of $C_z$ for $|\bar{\psi}_+^R(z)\rangle$ to recover the initial state, and the terminal state has a phase difference of $\pi$. Therefore, the gauge-invariant Berry phase is quantized to $\pi$ after two cycles, implying a fractional VWN, $\mathcal{W}_\psi = \frac{\theta}{n\pi} = \frac{1}{2}$. The result can also be inferred from the phase rigidity. In Eq. (11), in the neighborhood of $\delta z \sim 0$, $|\bar{\psi}_{+,-}^R\rangle$ split following the factor $\sqrt{\delta z}$, and the pre-factor $\delta z^{-1/4}$ indicates that a $4\pi$-change in the argument of $\delta z$ (two cycles) generates a phase factor of $\pi$, which is the Berry phase.

The Berry phase associated with order-2 EPs was experimentally confirmed . in a microwave experiment[11,62], and has subsequently been observed in an optical billiard[63] and acoustic experiments[43]. All these experiments have used a "stroboscopic" approach, in which systems are constructed at various parametric points along a $C_z$ and then measured separately, and the Berry phase is either inferred from the mode profiles or extracted by imposing parallel transport in data processing. In Hermitian systems, the Berry phase is also induced by enforcing dynamic adiabatic pumping along $C_z$. However, dynamic evolution is very subtle in non-Hermitian systems because of the prone to non-adiabatic transitions (Box 1).

**EP geometry and higher-order EP.** An order-2 EP occurs when both the real and imaginary parts of a Hamiltonian's discriminant are equal to zero, and thus at least two degrees of freedom (DOFs) are required to find an EP solution. Hence, in the absence of any prior conditions, order-2 EPs may form continuous trajectories in a three-dimensional (3D) parameter space. Here, we denote a closed and unknotted EP line homeomorphic to $S^1$ as an exceptional ring, otherwise we denote it an exceptional arc (EA). If a system possesses symmetries such that the real or imaginary part of the discriminant is pinned at zero, then its EPs can form a surface (line) in a 3D (2D) space[64,65]. For example, a Dirac cone with an added imaginary mass term still satisfies pseudo-Hermiticity, which guarantees the realness of the discriminant, so an exceptional ring is spawned[66] in the 2D reciprocal space. However, if the non-Hermitian terms do not exhibit antilinear symmetry, only two isolated order-2 EPs are spawned[67]. By exploring a system's DOFs and (or) imposing additional symmetries, order-2 EPs can form surfaces[65], such as an EP torus[68], an EP saddle[69], and an EP sphere[70]. EP lines emerging in Bloch bands give rise to a kaleidoscope of EP-line semi-metals, such as rings[66,71–73], knots[44,74], links and chain[75–77]. Recent theoretical studies have also discovered non-Hermitian higher-order Dirac and Weyl semimetals[78–80], enabling rich phenomena such as exceptional Fermi rings connected by hinge arcs, non-Hermitian Weyl nodes. In practical terms, EP geometries offer two advantages over a stand-alone EP. First, they alleviate the stringent parameter requirements for accessing isolated EPs, which in turn increases the robustness of EP-related phenomena[81–83]. Second, they can introduce anisotropic dependence of different parameters in eigenvalue and eigenvector splitting[43,56], which may be beneficial for applications such as EP-sensing with multiple sensitivities.

The variety of EP lines appearing in a space with dimensionality larger than two is abundant. However, they invariably entail the existence of loops belonging to distinct homotopy equivalence classes, which results in a nontrivial fundamental group for $\mathcal{M}_E(z)$. Therefore, each EP line and its generator from the corresponding eigenvalues and eigenvectors must be investigated to determine the topology of EP geometry[84,43], which we will demonstrate by revisiting Eq. (3). Here, we treat the hopping term $t$ in Eq. (3) as the third dimension. The two EPs in Fig. 1d evolve into two linearly crossing EAs shown in Fig. 2a, and their intersection at $(z, t) = (0, 0)$ is a diabolic point. Such a point belongs to a non-defective EP, a straightforward extension of the Hermitian degeneracy point[85,86]. The color and vector plots in the bottom plane respectively show the norm and phase gradient of the discriminant evaluated at $t = 1$, where the dashed purple loop lies. The norm vanishes at $z = \pm 2i$ and the phase gradient shows the same vorticity at both EPs. An inspection of the eigenvalue trajectories reveals that they braid into a Hopf link, and that each band forms an individual point gap (Fig. 2b); thus the net EWN is $\mathcal{W}_E = -2$, which is consistent with the result from Eq. (5). We can use the discriminant vorticity and the right-hand rule to assign a direction for the EAs, as indicated by the red arrow in Fig. 2a. The two EAs oriented in the same direction when viewed from the magenta plane. The oriented EAs can be analogously viewed as current-carrying wires and the discriminant field (phase gradient of the discriminant) being the magnetic field[74,77]. This reveals that there is a quantized net "current" flowing through the purple loop. A different loop (blue in Fig. 2a) can be drawn on the $xt$-plane at $y = -2$, and is nonhomotopic to the purple one, for they cannot transform into each other without intersecting an EA. The two EAs threading through the blue loop obviously have opposite directions such that the net "current" vanishes and the net EWN is $\mathcal{W}_E = 0$. These results confirm that the two loops belong to different equivalence classes and that the fundamental group of $\mathcal{M}_E$ is nontrivial. It is noteworthy that the topology of $\mathcal{M}_E$ bestows the seemingly trivial diabolic point at $(z, t) = (0, 0)$ with nontrivial topological characteristics, even though the system's two states are not even coupled ($t = 0$). The conditions for non-defective EPs to stably appear are analyzed in refs. [85,86].

The EP geometry becomes more interesting if a more sophisticated set of parameters is introduced. For example, we can expand Eq. (3) near the EP at $(z, t) = (-2i, 1)$ as follows

$$H_2 = \begin{bmatrix} 0 & 1-\delta t^2 \\ 1-\delta t^2 & -2i(1+\delta z) \end{bmatrix}, \qquad (17)$$

where $\delta t \in \mathbb{R}$ and $\delta z = \delta x + i\delta y \in \mathbb{C}$. The quadratic dependence of hopping on $\delta t$ can physically be realized by exploring the symmetry of wavefunctions[43,56]. The condition for EPs is $\delta t^2 + \delta z = 0$, which gives rise to a parabolic EA (Fig. 2c). Similarly, the direction of an EA is determined by the discriminant vorticity. Obviously, at least two nonhomotopic loops can be drawn, as shown in Fig. 2c. The EA threads through the blue loop once and hence it is a nontrivial loop. In contrast, the EA goes through the purple loop twice in opposite directions, so the loop is equivalently trivial. Indeed, translating the loop towards the positive $\delta t$ direction and it smoothly slides out of the parabola. From the discriminant vorticity, the EWNs of two intersecting EPs are opposite, so the net EWN is zero. In addition, the eigenvalue trajectories do not enclose any spectral area but form two lines in the complex energy plane, as shown in Fig. 2d. However, if we perform analytic continuation on $\delta t$ while enforcing $\delta z \in \mathbb{R}$, the resulting EAs are two parabolas osculating at the origin, with their two arms having the same direction, as shown in Fig. 2e. The eigenvalue trajectories not only enclose spectral areas but braid into a Hopf link (Fig. 2f), implying that the osculating point is an EP with $\mathcal{W}_E = -2$. Note that this loop can smoothly slide up and down without intersecting any EAs.

A multi-band Hamiltonian affords much greater possibilities in the EP geometry. Particularly, more than two bands can coalesce at the same points to form higher-order EPs, which have drawn considerable attention in the past few years. For example, order-3 EPs have been experimentally realized in photonic micro-ring resonators[87] and acoustic metasurfaces[88]. Order-4 EPs have been experimentally realized with acoustic cavities[57], microwave resonators[89], and in a single-photon interferometer[90]. An EP of order-6 was also achieved in electronic circuits[91]. Recipes to achieve EPs of an arbitrary order have been developed from various viewpoints, including supersymmetry, hidden symmetry and Jordan block form, amongst others[50,89,92–94] At an order-$N$ EP, the eigenvalue splitting typically follows $\delta z^{1/N}$, which yields a higher sensitivity enhancement compared to order-2 EPs[87,91].

The emergence of higher-order EPs can also be understood from a topological consideration[95,96] (Fig. 3). An order-$N$ EP can only form in an $M$-state Hamiltonian with $M \geq N$, which implies that the choices of parameters increase rapidly. It has been derived that, in the absence of any symmetry, an order-$N$ EP generally requires the tuning of $2(N-1)$ real parameters[97–99]. We illustrate some basics by considering the following three-state system[100]

$$H_3(\Lambda, \Xi) = \begin{bmatrix} \sqrt{2}i(1+\Lambda) & -1 & 0 \\ -1 & i\Xi & -1 \\ 0 & -1 & -\sqrt{2}i(1+\Lambda) \end{bmatrix}. \qquad (18)$$

Generically, without symmetry, an order-3 EP can emerge in a four-dimensional (4D) parameter space. This condition is met by letting both $\Lambda$ and $\Xi$ be complex parameters. The discriminant calculation indicates that the order-2 EPs form 2D surfaces, which generically intersect at isolated points in the 4D ambient space. To conveniently identify these points, we restrict to 3D cuts, and the EP surfaces become 1D EAs in the 3D subspace. An example in the $\Lambda_i \Xi$-space with $\Lambda_r = 0$ is shown in Fig. 3a. Making a further cut at $\Lambda_i = -0.2$ gives the eigenvalue Riemann surfaces shown in Fig. 3b, wherein two order-2 EPs are formed by different pairs of bands. This implies that the manner in which the blue and the red EAs in Fig. 3a intersect distinguishes the cases in a two-state system. Indeed, the two EAs coalesce at $\Lambda = 0$ (Fig. 3a) and all three eigenvalue sheets touch at a single point ($\Xi = 0$), generating an order-3 EP (Fig. 3c). Such a higher-order EP distinguishes from order-2 EPs in the following aspects. First, a path encircling the order-3 EP traverses all three Riemann sheets to form a closed loop, and the resultant eigenvalue braids have three strands that together form a single point gap (Fig. 3h). It follows that the parallel transport of a state must also cover three cycles around the order-3 EP to restore itself. The Berry phase accumulated in the process is $2\pi$, as shown in the top panel of Fig. 3g. As in an order-2 EP, this fractional winding behavior roots in the denominator of the eigenvector, and thus can be determined from the critical exponents of phase rigidity. The blue and magenta lines in Fig. 3f are the double-log plot of the phase rigidity and the detuning parameter near an order-2 and an order-3 EP, respectively. Their slopes, namely the critical exponents, are 1/2 and 2/3 respectively, which reveals a fundamental difference between them. It has been proved that the exponent is generally $(N-1)/N$ for an order-$N$ EP[95,96].

The rich topological characteristics of the order-3 EP are revealed by interrogating it in a different subspace. For example, in Fig. 3d we plot the eigenvalue Riemann surfaces in the $\Lambda$-plane with $\Theta = 0$. At first sight, it appears that the middle state decouples from the other two, which is not true because the phase rigidity of the middle state also vanishes at $\Lambda = 0$, as seen in the colormap in Fig. 3d. Figure 3i shows the eigenvalue braids around the order-3 EP, in which the middle eigenvalue (black) actually is linked with the other two, forming a Hopf link. This is a typical feature for an order-2 EP, and it implies that the upper/bottom (middle) state can recover after only two (one) cycles around the order-3 EP, which is clearly different from the situation in Fig. 3h. The Berry phases accumulated in this process are $2\pi$ ($\pi$) for the upper/bottom (middle) state, as shown in the bottom panel of Fig. 3g. The critical exponent of phase rigidity is unity in the $\Lambda$-plane (the black line in Fig. 3f), which follows a different $(N-1)/2$ law for an order-$N$ EP[50,89,96]. This dramatic difference between the $\Lambda$ and $\Theta$ planes highlights the unique hybrid nature of higher-order EPs, which is absent for order-2 EPs.

Because the evolution of non-Hermitian states is smoothly connected by spectral topology, multi-state non-Hermitian systems are an excellent platform for realizing non-Abelian permutations of states[101–103]. For example, the parallel transport around the blue (red) EA in Fig. 3a swaps states 2 and 3 (states 1 and 2), which is the consequence of a three-state unitary transformation captured by an SO(3) group. These two operations can generate all possible state permutations in a three-state system, which map to the dihedral group of order 3 – the smallest non-Abelian group[103], whose characteristics are attained by concatenating the generating operations in different orders, as shown in Fig. 3e. The non-Abelian permutation also fundamentally distinguishes the possible topology of a two-statl non-Hermitian system from a multi-state one.

**Non-Hermitian band topology and skin effects.** Band topology is one of the most successful applications of topology in physics. The most salient feature of topological matter is the existence of topological boundary modes (TBMs). The bulk-boundary correspondence states that every topological bulk state has its characteristic TBMs[104]. As spectral topology plays no role in Hermitian systems, band topology in Hermitian contexts only concerns the twisting of eigenvectors (wavefunctions) of the bulk states. In contrast, the complex eigenvalues existing in non-Hermitian systems give rise to spectral topology that also emerges in non-Hermitian periodic systems, manifested as the winding of bands driven by crystal momentum. As such, band topology in non-Hermitian contexts expands to include both spectral and wavefunction topology. However, these two layers of topology were initially regarded as distinct aspects that do not interfere with each other. For example, in ref. [32], the Chern number, an invariant of wavefunction topology that can be computed from VWN (for example, the Zak phase), was shown to be well-defined for a general non-Hermitian band, irrespective of the spectral topology. Another example in point is the Weyl exceptional ring, which simultaneously exhibits the wavefunction topology of a Weyl point by carrying a non-zero topological charge, and the spectral topology of order-2 EPs[73,105].

The discovery of the NHSEs revealed that a given wavefunction was not invariably independent of spectral topology. Several studies by different groups have shown that for certain non-Hermitian Hamiltonians[106–109], TBMs are found under open-boundary condition (OBC), but the topological invariant is ill-defined under periodic-boundary condition (PBC) due to the presence of EPs. This apparent violation of the bulk-boundary correspondence is attributed to the NHSE, which has two implications. First, the spectra under the OBC are drastically different from those under a PBC. Second, a large number of "skin modes" appear under the OBC. A skin mode is an eigenstate under the OBC that localizes at one boundary, and hence is not a superposition of Bloch wavefunctions. Although a skin mode resembles a TBM in terms of the wavefunction, the number of skin modes (TBMs) scales with the size of the bulk (boundary) in real space. Due to the NHSE, most OBC eigenstates become skin modes that differ drastically from Bloch wavefunctions under a PBC. It is therefore unsurprising that, the topological invariant, conventionally defined only for Bloch wavefunctions, fails to predict the wavefunction topology and the TBMs thereof under OBC. The NHSE has been realized experimentally in optics[110], electrical circuits[111,112], quantum walks[113], cold atoms[114], and other classical wave systems[115–117].

We can use a 1D Su–Schrieffer–Heeger (SSH) model with nonreciprocal hopping[107,118] (Fig. 4a) to illustrate the NHSE. The PBC Hamiltonian is

$$H(k) = \begin{bmatrix} 0 & \left(t_1 + \frac{\gamma}{2}\right) + t_2 e^{-ik} \\ \left(t_1 - \frac{\gamma}{2}\right) + t_2 e^{ik} & 0 \end{bmatrix}, \tag{15}$$

where $t_1$ and $t_2$ are respectively the intercell and intracell hopping coefficients. Here, $t_1$ is modified by the terms $\pm \gamma/2$, which indicates that the hopping towards the left is different from that toward the right. For simplicity, we set $t_1, t_2, \gamma \in \mathbb{R}$ with all three being positive, and $t_1 - \frac{\gamma}{2} > 0$. The terms $\pm \gamma/2$ generate nonreciprocal hopping that breaks the Hermiticity. The energy spectrum of Eq. (15) is shown in Fig. 4b. The gap-closing points can be identified at $t_1 = t_2 \pm \gamma/2$. Hermitian topological band theory indicates there are topological phase transitions at these gap-closing parameters. However, this prediction is inconsistent with the results from the OBC Hamiltonian,

$$\mathcal{H} = \sum_{n=1}^{N} \left( t_1 a_{n,A}^\dagger a_{n,B} + t_2 a_{n,B}^\dagger a_{n-1,A} + \text{h.c.} \right) + \sum_{n=1}^{N} \frac{\gamma}{2} \left( a_{n,A}^\dagger a_{n,B} - a_{n,B}^\dagger a_{n,A} \right), \tag{16}$$

where $a_{n,A(B)}^\dagger$ and $a_{n,A(B)}$ are the creation and annihilation operators in the $n^{\text{th}}$ unit cell on sublattice A(B), as shown in Fig. 4a. The OBC spectrum of Eq. (16) deviates drastically from the PBC spectrum (Fig. 4b). In particular, the gap-closing point of Eq. (16) does not align with the prediction of Eq. (15). Instead, the topological phase transition, which marked by the appearance or disappearance of topological zero modes, occur at gap-closing points in the OBC spectrum instead of PBC spectrum. A calculation of the eigenstates under the OBC reveals that they are almost all skin modes localized on the left edge. This model therefore exhibits the NHSE.

There is a simple explanation for the NHSE in this model. Inside the bulk, for $\gamma > 0$, the hopping amplitude is larger toward the left and smaller toward the right, thus any wave packet experiences a leftward push. Under a PBC, this tendency creates a loop current, such that plane-wave solutions preserve their forms and remain eigenstates. In contrast, under the OBC, this tendency collapses the wave packets toward the left until they hit a "hard wall," in this case, the left boundary.

This would typically result in reflection by the boundary; however, this reflection is suppressed by the weak rightward hopping. The OBC wavefunctions thus become skin modes that build up at the left boundary (Fig. 4c), which is a manifestation of the NHSE. Such a congregation of modes is possible because, as mentioned above, non-Hermitian eigenvectors are mutually skewed.

A closer inspection of this simple example reveals the physical origin of the NHSE: the failure of standing-wave formation. To understanding this, one may first inquire why the Bloch theorem works so well for predicting the bulk spectrum in almost all Hermitian bands, even under the OBC. One mathematically precise but somewhat technical answer is the Cauchy's interlacing theorem[119]. Here, we provide an alternative explanation based on physical arguments and using a 1D example.

Consider a single Hermitian band $E(k) \in \mathbb{R}$, where due to the periodicity $E(k) = E(k + 2\pi)$, for any generic $k_1 \in [0, 2\pi)$, there is another $k_2$, such that $E(k_1) = E(k_2)$. This is shown in Fig. 4d (left). Let $\psi_{1,2}(x) = e^{ik_{1,2}x}$ be two PBC eigenstates, where $k_i = 2\pi n_i/L$, $n_i \in \mathbb{Z}$, and $L$ is the length of the system. Neither of these PBC eigenstates satisfies the OBC $\psi(x = 0) = \psi(x = L) = 0$; however, a linear combination $\psi = -i(\psi_1 - \psi_2)/2 = e^{i(k_1+k_2)x/2} \sin[(k_1 - k_2)x/2] = \exp[i(n_1 + n_2)x\pi/L] \sin[(n_1 - n_2)x\pi/L]$ does. Moreover, because $E(k_1) = E(k_2)$, $\psi$ is also an eigenstate of energy $E(k_1)$. This means that, in Hermitian systems, two or more plane waves of the same energy can generically form standing waves that satisfy the OBC. (There are exceptions for band tops and bottoms, where the dispersion relation is an extremum. But the number of these points is of order $\mathcal{O}(1)$.) This is the reason why PBC eigenstates can effectively predict OBC wavefunctions, and why their spectra converge. However, the same argument fails for non-Hermitian Hamiltonians at the first step: the eigenvalues are complex and each band generically forms a point gap. So for a generic $k_1$, there is no $k_2$ such that $E(k_1) = E(k_2)$, as plotted in Fig. 4d (middle). In our model [Eq. (15)], the PBC spectrum in Fig. 4b forms a loop, such that the mapping from $k$ to $E(k)$ is one-to-one. A single plane wave cannot form standing-wave solution satisfying OBC, so the OBC and the PBC spectra differ (Fig. 4b). This explanation also accounts for the reason why, in some cases, the NHSE is absent even when the system is non-Hermitian in character: it remains possible to find $E(k_1) = E(k_2)$ if a complex-value band encloses zero spectral area. $\mathcal{P}$-symmetric 1D lattices belong to this case, and so do $\mathcal{T}^\dagger$-symmetric ones, as these symmetries ensure $E(k) = E(-k)$ (Fig. 4d, right).

Now, we are one step away from establishing a qualitative relationship between the spectral topology and the NHSE. First, if any part of the PBC spectrum forms a loop on the complex plane, then that loop cannot be part of the OBC spectrum because standing waves cannot form; thus the NHSE occurs. Conversely, if the PBC spectrum collapses into an arc, then there must be at least two momenta for every point on the spectrum, from which a standing wave solution is recovered, and hence the absence of the NHSE. The observation is directly connected to the spectral topology of the PBC spectrum. If the PBC spectrum forms a loop, its EWN with respect to any interior energy must be nonzero. In contrast, if it encloses no spectral area, the EWN must vanish. We can thereby conclude that the NHSE occurs if and only if the EWN of the PBC band is nonzero[120,121]. In higher-than-one dimensions, a similar statement can be obtained[122]. These results reveal the topological origin of the NHSE. Meanwhile, the above picture also explains the possible existence of extended bulk modes even in the presence of NHSE at isolated energies denoted Bloch points[123,124], at which two PBC spectral loops cross. Apparently, at a Bloch point, it remains feasible to construct an OBC eigenstate from two PBC eigenstates sharing the same energy.

While the above shows that the existence of NHSE is related to the spectral topology, a quantitative description of the skin modes is still needed for some applications, such as the re-establishment of bulk-boundary correspondence and the calculation of the topological invariant for non-Hermitian bands under OBC. The methods to this end include the non-Bloch band theory approach that relies on constructing a generalized Brillouin zone (GBZ) [107,118,125–127], or by exploiting a real-space topological invariant such as the biorthogonal polarization[109,128], the open-bulk winding number[123], or by using a doubled Hamiltonian[129]. For 1D systems, a general recipe, known as the GBZ method[107,118], has been extensively used. This method is mathematically related to the "banded Toeplitz matrices"[119]. Below, we demonstrate the non-Bloch band theory by establishing a GBZ using the model in Eqs. (15, 16).

One first observes that $\mathcal{H}$ (Eq. 16) is a pseudo-Hermitian matrix with an entirely real spectrum, which permits a similarity transformation to a Hermitian matrix,

$$\bar{\mathcal{H}} = S^{-1}\mathcal{H}S, \qquad (17)$$

where $S = \text{diag}(1, r, r, r^2, r^2 \dots r^{N-1}, r^{N-1}, r^N)$ with $r = \sqrt{\left|\left(t_1 - \frac{\gamma}{2}\right)/\left(t_1 + \frac{\gamma}{2}\right)\right|}$. $\bar{\mathcal{H}}$ describes a Hermitian SSH model with a set of modified hopping parameters $\bar{t}_1 = \sqrt{\left(t_1 - \frac{\gamma}{2}\right)\left(t_1 + \frac{\gamma}{2}\right)}$, and $\bar{t}_2 = t_2$. From these modified hoppings, we can work out a new topological transition point, $\bar{t}_1 = \bar{t}_2$, which is different from the gap-closing point of Eq. (15). This method only works for the OBC but not for the PBC, as it breaks the translation symmetry. A bulk eigenstate of the Hermitian $\bar{\mathcal{H}}$ is simply a Bloch wave $e^{ikx}$, upon which $S^{-1}$ acts to yield an eigenstate of $\mathcal{H}$, namely $r^{-x}e^{ikx} = e^{i[k+i\log(r)]x} = e^{ik'x}$, where $k' \equiv k + i\log(r)$ is a complex "momentum" that covers $i\log(r)$ to $2\pi + i\log(r)$ for $k \in [0, 2\pi)$. The imaginary part of $k'$ can be effectively treated as an imaginary magnetic flux[130,131]. It is

also related to a duality between non-Hermitian models in flat space and Hermitian ones in curved space[132]. Replacing $k$ with $k'$ in the dispersion relation under a PBC generates the spectrum of the OBC Hamiltonian $\bar{\mathcal{H}}$. Therefore, by shifting the Brillouin zone (BZ) by $i\log(r)$ on the complex energy plane, the OBC spectrum is obtained from the PBC spectrum, and the skin-mode wavefunctions are obtained from the Bloch wavefunctions. This shifted BZ of crystal momentum is the GBZ, as shown in Fig. 4e. In practice, it is more convenient to consider the exponential of $k'$, $\beta \equiv e^{ik'}$ such that the BZ is mapped to the unit circle $|\beta| = 1$, while the GBZ is mapped to $|\beta| = r$, a circle with a radius $r > 1$. In non-Hermitian studies, the BZ or GBZ usually refers to the range of $\beta$ instead of the range of $k$.

Not all non-Hermitian Hamiltonians under the OBC can be converted to Hermitian ones by similarity transformations, and neither are the GBZs always circular, especially in the presence of long-distance hopping[118,124]. One example is shown in Fig. 4e (right). It can be shown that the GBZ always exists for non-Hermitian bands in 1D, and it consists of loops enclosing the origin on the complex plane, where the number of loops equals that of bands. Refs. [126,127] show that finding the GBZ is equivalent to finding the roots of a complex polynomial, the degree of which does not scale with the size of a system. Once the GBZ is found, both the OBC spectrum and the eigenstates in the thermodynamic limit can be obtained. This is achieved by first solving the PBC Hamiltonian using the Bloch theorem and then replacing the momenta on the BZ with those on the GBZ. For a system of size $L$, the solution of a PBC Hamiltonian has the complexity $\mathcal{O}(L)$, because there are $L$ momenta; and computing the GBZ has complexity $\mathcal{O}(1)$. The total complexity is $\mathcal{O}(L)$, which is significantly lower than that of the numerical diagonalization, which is $\mathcal{O}(L^3)$.

The GBZ approach can also be extended for 2D systems, such as non-Hermitian Chern insulators[125,133] and has led to the discovery of corner skin modes[134–137]. However, a generic way to obtain GBZ in two or higher dimensions remains to be established. Nevertheless, by engineering the non-Hermitian DOFs, sophisticated versions of the NHSE have led to the discovery of new phenomena such as delocalization of topological modes[138–141], higher-order NHSE[142], hybrid skin-topological modes[134,135], and the anomalous scaling of coupled dissimilar skin modes[143]. NHSE also induces new types of robust dynamics, such as self-healing against scattering by spatial-temporal defects[144], edge burst[145], anharmonic Rabi oscillations[146], quantized response across the boundaries in a finite-size closed lattice with a specific boundary hopping[147]. A connection between NHSE and classical electrostatics has also been established[148].

Although the wavefunctions of skin modes accumulate toward an open boundary, their biorthogonal inner product remains unity [Eq. (9)]. This biorthogonality mandates that the skin-mode wavefunction's exponential-decay characteristic is "cancelled" by its corresponding left eigenvector, This property has been explored by analyzing the delocalization characteristics of a topological zero modes in a non-Hermitian SSH model, thereby bypassing the consideration of the bulk eigenstates under a PBC[28,109,128]. Owing to the system's chiral symmetry of a non-Hermitian SSH system, a topological mode is always pinned at zero energy when the lattice has an odd number of sites. Upon topological transition of the bulk, the zero mode delocalizes as the line gap closes, and it re-localizes at the opposite open boundary. By employing the non-Hermitian biorthogonality, the interference of the boundary-localized skin modes is removed so that the transition point can be correctly identified in an open lattice. A real-space topological invariant, denoted biorthogonal polarization, can then be constructed to correctly characterize topological transition. This method exemplifies the essential role of biorthogonality in non-Hermitian systems. Furthermore, inverse participation ratios—a quantity originally defined to benchmark the localization of Hermitian modes[149]—can also be extended by the biorthogonality of the non-Hermitian eigenvectors[133,150]. The resultant biorthogonal inverse participation ratio can distinguish skin modes from boundary modes, which leads to the correct bulk-boundary correspondence[133].

**Classification of non-Hermitian topological bands.** Studies have shown that the inclusion of non-Hermiticity considerably enriched the topological classification established in Hermitian band topology[26,30,31,151]. Herein, the terms "topology" and "topological" refer to the wavefunction topology, not the spectral topology. Also, the discussion here is restricted to the bands under a PBC, and the non-Hermitian Hamiltonian $H(k)$ is a mapping from a $d$-dimensional BZ to a square matrix. This section presumes the knowledge on the topological classification of Hermitian bands, and particularly, the "tenfold way."[29,152,153]

In the tenfold way, the topological classifications of gapped band structures only depend on two factors: symmetry and dimension. While "dimension" means the spatial dimension in which the system is supported in Hermitian and non-Hermitian cases, "symmetry" has new definitions. In Hermitian systems, onsite symmetries, the operations of which do not involve spatial DOFs, are classified into ten Altland-Zirnbauer (AZ) classes[154], labeled by their respective indices for the time-reversal symmetry (TRS), the particle-hole symmetry (PHS), and the chiral symmetry (CS). They are written as, respectively,

$$\mathcal{T}^\dagger H^* \mathcal{T} = H, \qquad (18)$$

$$\mathcal{C}^\dagger H^T \mathcal{C} = -H,$$
$$\mathcal{S}^{-1} H \mathcal{S} = -H,$$

where $\mathcal{T}$, $\mathcal{C}$, and $\mathcal{S}$ are unitary matrices for the time-reversal, the particle-hole, and the sublattice symmetry, respectively, written in the same single-particle basis under which the Hamiltonian is written. If $H$ is Hermitian, there are $H = H^\dagger$ and $H^* = H^T$, and Eqs. (18) become

$$\mathcal{T}^\dagger H^T \mathcal{T} = H,$$
$$\mathcal{C}^\dagger H^* \mathcal{C} = -H, \quad (19)$$
$$\mathcal{S}^{-1} H^\dagger \mathcal{S} = -H.$$

However, Eqs. (18) and (19) are no longer equivalent to each other when $H$ is non-Hermitian. This is called the "symmetry ramification", which means that one symmetry splits to two different symmetries. The symmetries defined in Eqs. (18) retain the names of TRS, PHS, and CS, while those defined in Eqs. (19) are customarily called TRS$^\dagger$, PHS$^\dagger$, and CS$^\dagger$. The symmetry classes defined by Eqs. (18) retain the name AZ classes, whereas those by Eqs. (19) acquire the name AZ$^\dagger$ classes[151]. Besides ramification, in non-Hermitian Hamiltonians, two distinct AZ classes may become one class. For example, if a Hamiltonian $H$ has TRS, then $iH$ has PHS$^\dagger$. The two classes containing these two symmetries, for example, class AI and class D$^\dagger$, are considered the same: this is called the "unification"[151,155,156]. After ramification and fusion, the original ten classes become 38 distinct classes[151,156].

Aside from symmetry, another key element in the topological classification is the definition of "gap." In Hermitian systems, a gap means a separation in energy between a group of bands (presumably the object of interest) and all the other bands in the system. In condensed matter, the Fermi energy gives a natural partition of bands into the bands below and the bands above that energy. In practice, the bands below are of particular interest, because the many-body ground state is made from filling out these bands. In general, a "gap" simply partitions all bands into two sets that have zero overlap in energy. This definition can be easily extended to non-Hermitian bands, which is now called the "line gap"[26,30,151], as shown in Fig. 1(a).

Consider a non-Hermitian band system under PBC, $\{E_{i=1,\ldots,N}(\boldsymbol{k}), \boldsymbol{k} \in \text{BZ}\}$. Each band $E_i(\boldsymbol{k})$ can be understood as a continuous mapping from BZ to the complex plane, the image of which is a loop for 1D, and a region of finite area in 2D and higher. Now consider a curve on the complex plane that is invariant under all symmetries of $H$, starts and ends at $\infty$, and does not intersect with any band or with itself. Such a curve necessarily gives a partition of the bands, dividing them into bands on one side, and bands on the other side, of this curve. This partition, or equivalently, this curve, is called a line gap. Refs. [30,151] show that as far as there is a line gap, the Hamiltonian can be made Hermitian following a continuous path that does not touch the line gap, and does not break any symmetry. After this step, the topological classification and diagnosis of a non-Hermitian Hamiltonian become those of a Hermitian Hamiltonian, well understood in the tenfold way.

As the idea of this "dual partition" and "band flattening" (a technique used in "Hermitianization") are directly borrowed from topological band theory for Hermitian bands, the topology of a line gap belongs to the wavefunction topology. The bulk-boundary correspondence is also similar: a $d$-dimensional bulk band has $(d-1)$-dimensional boundary modes which connect the two sets of bands separated by the line gap[30,151].

As discussed in the section "Complex eigenvalues and spectral topology," non-Hermitian bands can also form "point gaps"[30,120,151]. The implications of point gaps on the classifications of non-Hermitian Bloch bands are summarized in Box 2.

**Outlook.** The development of non-Hermitian physics has followed a rather unconventional path. It was first investigated by theorists as a mathematical physics topic and subsequently verified in classical-wave experiments. A wide variety of applications exploiting the properties of EPs were then identified and demonstrated before non-Hermitian topology started to attract the attention of condensed matter theorists. This is the joint outcome of the universality of non-Hermitian physics and the versatility of classical wave systems. To this date, classical waves and photonics remain the dominant platforms for experimental investigations of non-Hermitian topics, although platforms in other realms, such as cold atoms and quantum optics, are beginning to show potential. We expect classical waves will continue to play an important role in future studies of non-Hermitian systems and the topology therein. It should also be pointed out that non-Hermitian formalisms are also applicable to many other scenarios that use matrices, e.g., force analysis[157], scattering problems[158]. Yet the implication of non-Hermitian topology to these areas remain unexplored so far.

The fact that non-Hermiticity is often generated by the interaction of a system with DOFs lying outside the states of interest suggests that non-Hermitian theories can be an excellent tool to holistically investigate the role of the external DOFs. The prevailing theoretical methods within the Hermitian framework treat the external DOFs as perturbations or mean fields. The accurate treatments of the external DOFs should be reformed when the non-Hermiticity goes beyond perturbation or even the Hamiltonian framework, but the topology in the non-Hermitian physics still plays its role and thus requires both the control of internal DOFs and the engineering of interactions with the environment. Together these methods constitute a rich toolbox that should support new discoveries and new applications. Currently, many aspects of the topology and the physics of higher-order EPs and EP structures in higher-dimensional space remain under-explored. And the study of in non-Hermitian band topology is still in its infancy. For example, a general theory of NHSE in lattices with dimensions higher than one has yet to be devised. The interplay of band topology and EP topology in multi-band systems awaits systematic investigations. Furthermore, the interaction between NHSE and topological modes may generate topological phenomena inconceivable in Hermitian systems.

Non-Hermitian topology is currently a curiosity-driven topic, so it is unclear whether topological properties (like spectral winding and eigenvalue braids) and state permutations can be manifested in physical forms that have practical utility. Nevertheless, the non-Hermitian topology that occurs in the evolution of eigenvectors has been leveraged for useful applications such as waveguiding and metasurfaces[159]. We, therefore, anticipate that more exciting applications will continue to emerge as our understanding of non-Hermitian topology deepens and its connection with other realms of physics strengthens.

**References**

1. Scholtz, F. G., Geyer, H. B. & Nahne, F. J. W. Quasi-Hermitian Operators in Quantum Mechanics and the Variational Principle. *Ann. Phys.* **213**, 74–101 (1992).
2. Mostafazadeh, A. Pseudo-Hermiticity versus PT symmetry: The necessary condition for the reality of the spectrum of a non-Hermitian Hamiltonian. *J. Math. Phys.* **43**, 205–214 (2002).
3. Bender, C. M. & Boettcher, S. Real Spectra in Non-Hermitian Hamiltonians Having P T Symmetry. *Phys. Rev. Lett.* **80**, 5243–5246 (1998).
4. Bender, C. M., Brody, D. C. & Jones, H. F. Complex Extension of Quantum Mechanics. *Phys. Rev. Lett.* **89**, 270401 (2002).
5. Dorey, P., Dunning, C. & Tateo, R. Spectral equivalences, Bethe ansatz equations, and reality properties in PT -symmetric quantum mechanics. *J Phys Math Gen* **34**, 5679–5704 (2001).
6. Dorey, P., Dunning, C. & Tateo, R. The ODE/IM correspondence. *J. Phys. Math. Theor.* **40**, R205–R283 (2007).
7. Mostafazadeh, A. Pseudo-Hermiticity versus PT-symmetry II: A complete characterization of non-Hermitian Hamiltonians with a real spectrum. *J. Math. Phys.* **43**, 2814 (2002).
8. Mostafazadeh, A. Pseudo-Hermiticity versus PT-symmetry III: Equivalence of pseudo-Hermiticity and the presence of antilinear symmetries. *J. Math. Phys.* **43**, 3944–3951 (2002).
9. Bender, C. M. Making sense of non-Hermitian Hamiltonians. *Rep. Prog. Phys.* **70**, 947–1018 (2007).
10. Philipp, M., Brentano, P. von, Pascovici, G. & Richter, A. Frequency and width crossing of two interacting resonances in a microwave cavity. *Phys. Rev. E* **62**, 1922–1926 (2000).
11. Dembowski, C. *et al.* Experimental Observation of the Topological Structure of Exceptional Points. *Phys. Rev. Lett.* **86**, 787–790 (2001).
12. Stehmann, T., Heiss, W. D. & Scholtz, F. G. Observation of exceptional points in electronic circuits. *J. Phys. Math. Gen.* **37**, 7813–7819 (2004).

13. Guo, A. *et al.* Observation of P T -Symmetry Breaking in Complex Optical Potentials. *Phys. Rev. Lett.* **103**, (2009).
14. Rüter, C. E. *et al.* Observation of parity–time symmetry in optics. *Nat. Phys.* **6**, 192–195 (2010).
15. Choi, Y. *et al.* Quasieigenstate Coalescence in an Atom-Cavity Quantum Composite. *Phys. Rev. Lett.* **104**, 153601 (2010).
16. Bittner, S. *et al.* P T Symmetry and Spontaneous Symmetry Breaking in a Microwave Billiard. *Phys. Rev. Lett.* **108**, 024101 (2012).
17. Kato, T. *Perturbation theory for linear operators*. (Springer, 1995).
18. Heiss, W. D. & Sannino, A. L. Avoided level crossing and exceptional points. *J. Phys. Math. Gen.* **23**, 1167–1178 (1990).
19. Heiss, W. D. Phases of wave functions and level repulsion. *Eur. Phys. J. - At. Mol. Opt. Phys.* **7**, 1–4 (1999).
20. Longhi, S. Parity-time symmetry meets photonics: A new twist in non-Hermitian optics. *EPL Europhys. Lett.* **120**, 64001 (2017).
21. Feng, L., El-Ganainy, R. & Ge, L. Non-Hermitian photonics based on parity–time symmetry. *Nat. Photonics* **11**, 752–762 (2017).
22. El-Ganainy, R. *et al.* Non-Hermitian physics and PT symmetry. *Nat. Phys.* **14**, 11–19 (2018).
23. Miri, M.-A. & Alù, A. Exceptional points in optics and photonics. *Science* **363**, eaar7709 (2019).
24. Özdemir, Ş. K., Rotter, S., Nori, F. & Yang, L. Parity–time symmetry and exceptional points in photonics. *Nat. Mater.* **18**, 783–798 (2019).
25. Parto, M., Liu, Y. G. N., Bahari, B., Khajavikhan, M. & Christodoulides, D. N. Non-Hermitian and topological photonics: optics at an exceptional point. *Nanophotonics* **10**, 403–423 (2020).
26. Ashida, Y., Gong, Z. & Ueda, M. Non-Hermitian physics. *Adv. Phys.* **69**, 249–435 (2020).
27. Wang, H. *et al.* Topological physics of non-Hermitian optics and photonics: a review. *J. Opt.* **23**, 123001 (2021).
28. Bergholtz, E. J., Budich, J. C. & Kunst, F. K. Exceptional topology of non-Hermitian systems. *Rev. Mod. Phys.* **93**, 015005 (2021).
29. Chiu, C.-K., Teo, J. C. Y., Schnyder, A. P. & Ryu, S. Classification of topological quantum matter with symmetries. *Rev. Mod. Phys.* **88**, 035005 (2016).
30. Gong, Z. *et al.* Topological Phases of Non-Hermitian Systems. *Phys. Rev. X* **8**, 031079 (2018).
31. Kawabata, K., Bessho, T. & Sato, M. Classification of Exceptional Points and Non-Hermitian Topological Semimetals. *Phys. Rev. Lett.* **123**, 066405 (2019).
32. Shen, H., Zhen, B. & Fu, L. Topological Band Theory for Non-Hermitian Hamiltonians. *Phys. Rev. Lett.* **120**, 146402 (2018).
33. Wiersig, J. Enhancing the Sensitivity of Frequency and Energy Splitting Detection by Using Exceptional Points: Application to Microcavity Sensors for Single-Particle Detection. *Phys. Rev. Lett.* **112**, (2014).
34. Chen, W., Kaya Özdemir, Ş., Zhao, G., Wiersig, J. & Yang, L. Exceptional points enhance sensing in an optical microcavity. *Nature* **548**, 192–196 (2017).
35. Lai, Y.-H., Lu, Y.-K., Suh, M.-G., Yuan, Z. & Vahala, K. Observation of the exceptional-point-enhanced Sagnac effect. *Nature* **576**, 65–69 (2019).
36. Hokmabadi, M. P., Schumer, A., Christodoulides, D. N. & Khajavikhan, M. Non-Hermitian ring laser gyroscopes with enhanced Sagnac sensitivity. *Nature* **576**, 70–74 (2019).
37. Heiss, W. D. & Harney, H. L. The chirality of exceptional points. *Eur. Phys. J. D* **17**, 149–151 (2001).
38. Dembowski, C. *et al.* Observation of a Chiral State in a Microwave Cavity. *Phys. Rev. Lett.* **90**, (2003).
39. Heiss, W. D. The physics of exceptional points. *J. Phys. Math. Theor.* **45**, 444016 (2012).
40. Leykam, D., Bliokh, K. Y., Huang, C., Chong, Y. D. & Nori, F. Edge Modes, Degeneracies, and Topological Numbers in Non-Hermitian Systems. *Phys. Rev. Lett.* **118**, 040401 (2017).

41. Yang, Z., Schnyder, A. P., Hu, J. & Chiu, C.-K. Fermion Doubling Theorems in Two-Dimensional Non-Hermitian Systems for Fermi Points and Exceptional Points. *Phys. Rev. Lett.* **126**, 086401 (2021).
42. Wang, K. *et al.* Generating arbitrary topological windings of a non-Hermitian band. *Science* **371**, 1240–1245 (2021).
43. Tang, W., Ding, K. & Ma, G. Direct Measurement of Topological Properties of an Exceptional Parabola. *Phys. Rev. Lett.* **127**, 034301 (2021).
44. Carlström, J., Stålhammar, M., Budich, J. C. & Bergholtz, E. J. Knotted non-Hermitian metals. *Phys. Rev. B* **99**, 161115 (2019).
45. Hu, H. & Zhao, E. Knots and Non-Hermitian Bloch Bands. *Phys. Rev. Lett.* **126**, 010401 (2021).
46. Wang, K., Dutt, A., Wojcik, C. C. & Fan, S. Topological complex-energy braiding of non-Hermitian bands. *Nature* **598**, 59–64 (2021).
47. Patil, Y. S. S. *et al.* Measuring the knot of degeneracies and the eigenvalue braids near a third-order exceptional point. *arXiv:2112.00157* (2021).
48. Weigert, S. Completeness and orthonormality in PT-symmetric quantum systems. *Phys. Rev. A* **68**, 062111 (2003).
49. Brody, D. C. Biorthogonal quantum mechanics. *J. Phys. Math. Theor.* **47**, 035305 (2014).
50. Xiao, Y.-X., Zhang, Z.-Q., Hang, Z. H. & Chan, C. T. Anisotropic exceptional points of arbitrary order. *Phys. Rev. B* **99**, 241403 (2019).
51. Chen, H.-Z. *et al.* Revealing the missing dimension at an exceptional point. *Nat. Phys.* **16**, 571–578 (2020).
52. Lee, C. H. Exceptional Bound States and Negative Entanglement Entropy. *Phys. Rev. Lett.* **128**, 010402 (2022).
53. Chang, P.-Y., You, J.-S., Wen, X. & Ryu, S. Entanglement spectrum and entropy in topological non-Hermitian systems and nonunitary conformal field theory. *Phys. Rev. Res.* **2**, 033069 (2020).
54. Rotter, I. A non-Hermitian Hamilton operator and the physics of open quantum systems. *J. Phys. Math. Theor.* **42**, 153001 (2009).
55. Bulgakov, E. N., Rotter, I. & Sadreev, A. F. Phase rigidity and avoided level crossings in the complex energy plane. *Phys. Rev. E* **74**, 056204 (2006).
56. Ding, K., Ma, G., Zhang, Z. Q. & Chan, C. T. Experimental Demonstration of an Anisotropic Exceptional Point. *Phys. Rev. Lett.* **121**, 085702 (2018).
57. Ding, K., Ma, G., Xiao, M., Zhang, Z. Q. & Chan, C. T. Emergence, Coalescence, and Topological Properties of Multiple Exceptional Points and Their Experimental Realization. *Phys. Rev. X* **6**, 021007 (2016).
58. Wojcik, C. C., Sun, X.-Q., Bzdušek, T. & Fan, S. Homotopy characterization of non-Hermitian Hamiltonians. *Phys. Rev. B* **101**, 205417 (2020).
59. Li, Z. & Mong, R. S. K. Homotopical characterization of non-Hermitian band structures. *Phys. Rev. B* **103**, 155129 (2021).
60. Vanderbilt, D. *Berry Phases in Electronic Structure Theory: Electric Polarization, Orbital Magnetization and Topological Insulators*. (Cambridge University Press, 2018).
61. Lee, S.-Y., Ryu, J.-W., Kim, S. W. & Chung, Y. Geometric phase around multiple exceptional points. *Phys. Rev. A* **85**, 064103 (2012).
62. Dembowski, C. *et al.* Encircling an exceptional point. *Phys. Rev. E* **69**, 056216 (2004).
63. Lee, S.-B. *et al.* Observation of an Exceptional Point in a Chaotic Optical Microcavity. *Phys. Rev. Lett.* **103**, (2009).
64. Budich, J. C., Carlström, J., Kunst, F. K. & Bergholtz, E. J. Symmetry-protected nodal phases in non-Hermitian systems. *Phys. Rev. B* **99**, 041406 (2019).
65. Okugawa, R. & Yokoyama, T. Topological exceptional surfaces in non-Hermitian systems with parity-time and parity-particle-hole symmetries. *Phys. Rev. B* **99**, 041202 (2019).

66. Zhen, B. *et al.* Spawning rings of exceptional points out of Dirac cones. *Nature* **525**, 354–358 (2015).
67. Zhou, H. *et al.* Observation of bulk Fermi arc and polarization half charge from paired exceptional points. *Science* **359**, 1009–1012 (2018).
68. Zhou, H., Lee, J. Y., Liu, S. & Zhen, B. Exceptional surfaces in PT-symmetric non-Hermitian photonic systems. *Optica* **6**, 190 (2019).
69. Zhang, X., Ding, K., Zhou, X., Xu, J. & Jin, D. Experimental Observation of an Exceptional Surface in Synthetic Dimensions with Magnon Polaritons. *Phys. Rev. Lett.* **123**, 237202 (2019).
70. Rui, W. B., Hirschmann, M. M. & Schnyder, A. P. PT -symmetric non-Hermitian Dirac semimetals. *Phys. Rev. B* **100**, 245116 (2019).
71. Szameit, A., Rechtsman, M. C., Bahat-Treidel, O. & Segev, M. P T -symmetry in honeycomb photonic lattices. *Phys. Rev. A* **84**, 021806 (2011).
72. Cerjan, A., Raman, A. & Fan, S. Exceptional Contours and Band Structure Design in Parity-Time Symmetric Photonic Crystals. *Phys. Rev. Lett.* **116**, 203902 (2016).
73. Cerjan, A. *et al.* Experimental realization of a Weyl exceptional ring. *Nat. Photonics* **13**, 623–628 (2019).
74. Yang, Z., Chiu, C.-K., Fang, C. & Hu, J. Jones Polynomial and Knot Transitions in Hermitian and non-Hermitian Topological Semimetals. *Phys. Rev. Lett.* **124**, 186402 (2020).
75. Carlström, J. & Bergholtz, E. J. Exceptional links and twisted Fermi ribbons in non-Hermitian systems. *Phys. Rev. A* **98**, 042114 (2018).
76. Yang, Z. & Hu, J. Non-Hermitian Hopf-link exceptional line semimetals. *Phys. Rev. B* **99**, 081102 (2019).
77. Cui, X., Zhang, R.-Y., Chen, W.-J., Zhang, Z.-Q. & Chan, C. T. Symmetry-protected topological exceptional chains in non-Hermitian crystals. Preprint at http://arxiv.org/abs/2204.08052 (2022).
78. Ghorashi, S. A. A., Li, T., Sato, M. & Hughes, T. L. Non-Hermitian higher-order Dirac semimetals. *Phys. Rev. B* **104**, L161116 (2021).
79. Ghorashi, S. A. A., Li, T. & Sato, M. Non-Hermitian higher-order Weyl semimetals. *Phys. Rev. B* **104**, L161117 (2021).
80. Liu, T., He, J. J., Yang, Z. & Nori, F. Higher-Order Weyl-Exceptional-Ring Semimetals. *Phys. Rev. Lett.* **127**, 196801 (2021).
81. Zhong, Q. *et al.* Sensing with Exceptional Surfaces in Order to Combine Sensitivity with Robustness. *Phys. Rev. Lett.* **122**, 153902 (2019).
82. Qin, G. *et al.* Experimental Realization of Sensitivity Enhancement and Suppression with Exceptional Surfaces. *Laser Photonics Rev.* **15**, 2000569 (2021).
83. Soleymani, S. *et al.* Chiral and degenerate perfect absorption on exceptional surfaces. *Nat. Commun.* **13**, 599 (2022).
84. Tiwari, A. & Bzdušek, T. Non-Abelian topology of nodal-line rings in PT -symmetric systems. *Phys. Rev. B* **101**, 195130 (2020).
85. Xue, H., Wang, Q., Zhang, B. & Chong, Y. D. Non-Hermitian Dirac Cones. *Phys. Rev. Lett.* **124**, 236403 (2020).
86. Sayyad, S., Stalhammar, M., Rodland, L. & Kunst, F. K. Symmetry-protected exceptional and nodal points in non-Hermitian systems. Preprint at http://arxiv.org/abs/2204.13945 (2022).
87. Hodaei, H. *et al.* Enhanced sensitivity at higher-order exceptional points. *Nature* **548**, 187–191 (2017).
88. Fang, X. *et al.* Observation of higher-order exceptional points in a non-local acoustic metagrating. *Commun. Phys.* **4**, 271 (2021).
89. Wang, S. *et al.* Arbitrary order exceptional point induced by photonic spin–orbit interaction in coupled resonators. *Nat. Commun.* **10**, 832 (2019).
90. Bian, Z. *et al.* Conserved quantities in parity-time symmetric systems. *Phys. Rev. Res.* **2**, 022039 (2020).
91. Xiao, Z., Li, H., Kottos, T. & Alù, A. Enhanced Sensing and Nondegraded Thermal Noise Performance Based on P T -Symmetric Electronic Circuits with a Sixth-Order Exceptional Point. *Phys. Rev. Lett.* **123**, 213901 (2019).

92. Teimourpour, M. H., El-Ganainy, R., Eisfeld, A., Szameit, A. & Christodoulides, D. N. Light transport in PT-invariant photonic structures with hidden symmetries. *Phys. Rev. A* **90**, 053817 (2014).
93. Zhang, X. Z., Jin, L. & Song, Z. Perfect state transfer in PT-symmetric non-Hermitian networks. *Phys. Rev. A* **85**, 012106 (2012).
94. Zhang, S. M., Zhang, X. Z., Jin, L. & Song, Z. High-order exceptional points in supersymmetric arrays. *Phys. Rev. A* **101**, 033820 (2020).
95. Heiss, W. D. Chirality of wavefunctions for three coalescing levels. *J. Phys. Math. Theor.* **41**, 244010 (2008).
96. Demange, G. & Graefe, E.-M. Signatures of three coalescing eigenfunctions. *J. Phys. Math. Theor.* **45**, 025303 (2012).
97. Mandal, I. & Bergholtz, E. J. Symmetry and Higher-Order Exceptional Points. *Phys. Rev. Lett.* **127**, 186601 (2021).
98. Delplace, P., Yoshida, T. & Hatsugai, Y. Symmetry-Protected Multifold Exceptional Points and their Topological Characterization. *Phys. Rev. Lett.* **127**, 186602 (2021).
99. Sayyad, S. & Kunst, F. K. Realizing exceptional points of any order in the presence of symmetry. *arXiv:2202.07009* (2022).
100. Tang, W. *et al.* Exceptional nexus with a hybrid topological invariant. *Science* **370**, 1077–1080 (2020).
101. Schindler, S. T. & Bender, C. M. Winding in non-Hermitian systems. *J. Phys. Math. Theor.* **51**, 055201 (2018).
102. Zhong, Q., Khajavikhan, M., Christodoulides, D. N. & El-Ganainy, R. Winding around non-Hermitian singularities. *Nat. Commun.* **9**, 4808 (2018).
103. Tang, W., Ding, K. & Ma, G. Experimental Realization of Non-Abelian Permutations in a Three-State Non-Hermitian System. *Natl. Sci. Rev.* nwac010 (2022) doi:10.1093/nsr/nwac010.
104. Hasan, M. Z. & Kane, C. L. *Colloquium*: Topological insulators. *Rev. Mod. Phys.* **82**, 3045–3067 (2010).
105. Xu, Y., Wang, S.-T. & Duan, L.-M. Weyl Exceptional Rings in a Three-Dimensional Dissipative Cold Atomic Gas. *Phys. Rev. Lett.* **118**, 045701 (2017).
106. Martinez Alvarez, V. M., Barrios Vargas, J. E. & Foa Torres, L. E. F. Non-Hermitian robust edge states in one dimension: Anomalous localization and eigenspace condensation at exceptional points. *Phys. Rev. B* **97**, 121401 (2018).
107. Yao, S. & Wang, Z. Edge States and Topological Invariants of Non-Hermitian Systems. *Phys. Rev. Lett.* **121**, 086803 (2018).
108. McDonald, A., Pereg-Barnea, T. & Clerk, A. A. Phase-Dependent Chiral Transport and Effective Non-Hermitian Dynamics in a Bosonic Kitaev-Majorana Chain. *Phys. Rev. X* **8**, 041031 (2018).
109. Kunst, F. K., Edvardsson, E., Budich, J. C. & Bergholtz, E. J. Biorthogonal Bulk-Boundary Correspondence in Non-Hermitian Systems. *Phys. Rev. Lett.* **121**, 026808 (2018).
110. Weidemann, S. *et al.* Topological funneling of light. *Science* **368**, 311–314 (2020).
111. Helbig, T. *et al.* Generalized bulk–boundary correspondence in non-Hermitian topolectrical circuits. *Nat. Phys.* **16**, 747–750 (2020).
112. Hofmann, T. *et al.* Reciprocal skin effect and its realization in a topolectrical circuit. *Phys. Rev. Res.* **2**, 023265 (2020).
113. Xiao, L. *et al.* Non-Hermitian bulk–boundary correspondence in quantum dynamics. *Nat. Phys.* **16**, 761–766 (2020).
114. Liang, Q. *et al.* Observation of Non-Hermitian Skin Effect and Topology in Ultracold Atoms. *arXiv:2201.09478* (2022).
115. Brandenbourger, M., Locsin, X., Lerner, E. & Coulais, C. Non-reciprocal robotic metamaterials. *Nat. Commun.* **10**, 4608 (2019).

116. Ghatak, A., Brandenbourger, M., van Wezel, J. & Coulais, C. Observation of non-Hermitian topology and its bulk–edge correspondence in an active mechanical metamaterial. *Proc. Natl. Acad. Sci.* **117**, 29561–29568 (2020).
117. Zhang, L. *et al.* Acoustic non-Hermitian skin effect from twisted winding topology. *Nat. Commun.* **12**, 6297 (2021).
118. Yokomizo, K. & Murakami, S. Non-Bloch Band Theory of Non-Hermitian Systems. *Phys. Rev. Lett.* **123**, 066404 (2019).
119. Böttcher, A. & Grudsky, S. M. *Spectral properties of banded Toeplitz matrices*. (SIAM, 2005).
120. Okuma, N., Kawabata, K., Shiozaki, K. & Sato, M. Topological Origin of Non-Hermitian Skin Effects. *Phys. Rev. Lett.* **124**, 086801 (2020).
121. Zhang, K., Yang, Z. & Fang, C. Correspondence between Winding Numbers and Skin Modes in Non-Hermitian Systems. *Phys. Rev. Lett.* **125**, 126402 (2020).
122. Zhang, K., Yang, Z. & Fang, C. Universal non-Hermitian skin effect in two and higher dimensions. *Nat. Commun.* **13**, 2496 (2022).
123. Song, F., Yao, S. & Wang, Z. Non-Hermitian Topological Invariants in Real Space. *Phys. Rev. Lett.* **123**, 246801 (2019).
124. Longhi, S. Probing non-Hermitian skin effect and non-Bloch phase transitions. *Phys. Rev. Res.* **1**, 023013 (2019).
125. Yao, S., Song, F. & Wang, Z. Non-Hermitian Chern Bands. *Phys. Rev. Lett.* **121**, 136802 (2018).
126. Yang, Z., Zhang, K., Fang, C. & Hu, J. Non-Hermitian Bulk-Boundary Correspondence and Auxiliary Generalized Brillouin Zone Theory. *Phys. Rev. Lett.* **125**, 226402 (2020).
127. Lee, C. H., Li, L., Thomale, R. & Gong, J. Unraveling non-Hermitian pumping: Emergent spectral singularities and anomalous responses. *Phys. Rev. B* **102**, 085151 (2020).
128. Edvardsson, E., Kunst, F. K., Yoshida, T. & Bergholtz, E. J. Phase transitions and generalized biorthogonal polarization in non-Hermitian systems. *Phys. Rev. Res.* **2**, 043046 (2020).
129. Borgnia, D. S., Kruchkov, A. J. & Slager, R.-J. Non-Hermitian Boundary Modes and Topology. *Phys. Rev. Lett.* **124**, 056802 (2020).
130. Lee, C. H. & Thomale, R. Anatomy of skin modes and topology in non-Hermitian systems. *Phys. Rev. B* **99**, 201103 (2019).
131. Jin, L. & Song, Z. Bulk-boundary correspondence in a non-Hermitian system in one dimension with chiral inversion symmetry. *Phys. Rev. B* **99**, 081103 (2019).
132. Lv, C., Zhang, R., Zhai, Z. & Zhou, Q. Curving the space by non-Hermiticity. *Nat. Commun.* **13**, 2184 (2022).
133. Xiao, Y.-X. & Chan, C. T. Topology in non-Hermitian Chern insulator systems with skin effect. *ArXiv211102648 Cond-Mat* (2021).
134. Lee, C. H., Li, L. & Gong, J. Hybrid Higher-Order Skin-Topological Modes in Nonreciprocal Systems. *Phys. Rev. Lett.* **123**, 016805 (2019).
135. Zou, D. *et al.* Observation of hybrid higher-order skin-topological effect in non-Hermitian topolectrical circuits. *Nat. Commun.* **12**, 7201 (2021).
136. Zhang, X., Tian, Y., Jiang, J.-H., Lu, M.-H. & Chen, Y.-F. Observation of higher-order non-Hermitian skin effect. *Nat. Commun.* **12**, 5377 (2021).
137. Zhang, K., Yang, Z. & Fang, C. Universal non-Hermitian skin effect in two and higher dimensions. *ArXiv210205059 Cond-Mat* (2021).
138. Gao, P., Willatzen, M. & Christensen, J. Anomalous Topological Edge States in Non-Hermitian Piezophononic Media. *Phys. Rev. Lett.* **125**, 206402 (2020).
139. Zhu, W., Teo, W. X., Li, L. & Gong, J. Delocalization of topological edge states. *Phys. Rev. B* **103**, 195414 (2021).

140. Teo, W. X., Zhu, W. & Gong, J. Tunable two-dimensional laser arrays with zero-phase locking. *Phys. Rev. B* **105**, L201402 (2022).
141. Wang, W., Wang, X. & Ma, G. Non-Hermitian morphing of topological modes. *Nature* **608**, 50–55 (2022).
142. Kawabata, K. Higher-order non-Hermitian skin effect. *Phys. Rev. B* 16 (2020).
143. Li, L., Lee, C. H., Mu, S. & Gong, J. Critical non-Hermitian skin effect. *Nat. Commun.* **11**, 5491 (2020).
144. Longhi, S. Self-Healing of Non-Hermitian Topological Skin Modes. *Phys. Rev. Lett.* **128**, 157601 (2022).
145. Xue, W.-T., Hu, Y.-M., Song, F. & Wang, Z. Non-Hermitian Edge Burst. *Phys. Rev. Lett.* **128**, 120401 (2022).
146. Lee, C. H. & Longhi, S. Ultrafast and anharmonic Rabi oscillations between non-Bloch bands. *Commun. Phys.* **3**, 147 (2020).
147. Li, L., Mu, S., Lee, C. H. & Gong, J. Quantized classical response from spectral winding topology. *Nat. Commun.* **12**, 5294 (2021).
148. Yang, R. *et al.* Designing non-Hermitian real spectra through electrostatics. *arXiv:2201.04153* (2022).
149. Wegner, F. Inverse Participation Ratio in 2 + e Dimensions. *Z Phys B* **36**, 209 (1980).
150. Wang, P., Jin, L. & Song, Z. Non-Hermitian phase transition and eigenstate localization induced by asymmetric coupling. *Phys. Rev. A* **99**, 062112 (2019).
151. Kawabata, K., Shiozaki, K., Ueda, M. & Sato, M. Symmetry and Topology in Non-Hermitian Physics. *Phys. Rev. X* **9**, 041015 (2019).
152. Kitaev, A., Lebedev, V. & Feigel'man, M. Periodic table for topological insulators and superconductors. in *AIP Conference Proceedings* 22–30 (AIP, 2009). doi:10.1063/1.3149495.
153. Ryu, S., Schnyder, A. P., Furusaki, A. & Ludwig, A. W. W. Topological insulators and superconductors: tenfold way and dimensional hierarchy. *New J. Phys.* **12**, 065010 (2010).
154. Altland, A. & Zirnbauer, M. R. Nonstandard symmetry classes in mesoscopic normal-superconducting hybrid structures. *Phys. Rev. B* **55**, 1142–1161 (1997).
155. Kawabata, K., Higashikawa, S., Gong, Z., Ashida, Y. & Ueda, M. Topological unification of time-reversal and particle-hole symmetries in non-Hermitian physics. *Nat. Commun.* **10**, 297 (2019).
156. Zhou, H. & Lee, J. Y. Periodic table for topological bands with non-Hermitian symmetries. *Phys. Rev. B* **99**, 235112 (2019).
157. Li, X., Liu, Y., Lin, Z., Ng, J. & Chan, C. T. Non-Hermitian physics for optical manipulation uncovers inherent instability of large clusters. *Nat. Commun.* **12**, 6597 (2021).
158. Chong, Y. D., Ge, L. & Stone, A. D. P T -Symmetry Breaking and Laser-Absorber Modes in Optical Scattering Systems. *Phys. Rev. Lett.* **106**, (2011).
159. Song, Q., Odeh, M., Zúñiga-Pérez, J., Kanté, B. & Genevet, P. Plasmonic topological metasurface by encircling an exceptional point. *Science* **373**, 1133–1137 (2021).
160. Uzdin, R., Mailybaev, A. & Moiseyev, N. On the observability and asymmetry of adiabatic state flips generated by exceptional points. *J. Phys. Math. Theor.* **44**, 435302 (2011).
161. Berry, M. V. & Uzdin, R. Slow non-Hermitian cycling: exact solutions and the Stokes phenomenon. *J. Phys. Math. Theor.* **44**, 435303 (2011).
162. Milburn, T. J. *et al.* General description of quasiadiabatic dynamical phenomena near exceptional points. *Phys. Rev. A* **92**, 052124 (2015).
163. Hassan, A. U. *et al.* Chiral state conversion without encircling an exceptional point. *Phys. Rev. A* **96**, 052129 (2017).
164. Nasari, H. *et al.* Observation of chiral state transfer without encircling an exceptional point. *Nature* **605**, 256–261 (2022).

165. Gilary, I., Mailybaev, A. A. & Moiseyev, N. Time-asymmetric quantum-state-exchange mechanism. *Phys. Rev. A* **88**, 010102 (2013).
166. Doppler, J. *et al.* Dynamically encircling an exceptional point for asymmetric mode switching. *Nature* **537**, 76–79 (2016).
167. Xu, H., Mason, D., Jiang, L. & Harris, J. G. E. Topological energy transfer in an optomechanical system with exceptional points. *Nature* **537**, 80–83 (2016).
168. Hassan, A. U., Zhen, B., Soljačić, M., Khajavikhan, M. & Christodoulides, D. N. Dynamically Encircling Exceptional Points: Exact Evolution and Polarization State Conversion. *Phys. Rev. Lett.* **118**, 093002 (2017).
169. Zhang, X.-L., Wang, S., Hou, B. & Chan, C. T. Dynamically Encircling Exceptional Points: *In situ* Control of Encircling Loops and the Role of the Starting Point. *Phys. Rev. X* **8**, 021066 (2018).
170. Zhang, X.-L. & Chan, C. T. Hybrid exceptional point and its dynamical encircling in a two-state system. *Phys. Rev. A* **98**, (2018).
171. Yoon, J. W. *et al.* Time-asymmetric loop around an exceptional point over the full optical communications band. *Nature* **562**, 86–90 (2018).
172. Zhang, X.-L. & Chan, C. T. Dynamically encircling exceptional points in a three-mode waveguide system. *Commun. Phys.* **2**, 63 (2019).
173. Zhang, X.-L., Jiang, T. & Chan, C. T. Dynamically encircling an exceptional point in anti-parity-time symmetric systems: asymmetric mode switching for symmetry-broken modes. *Light Sci. Appl.* **8**, 88 (2019).
174. Geng, L., Zhang, W., Zhang, X. & Zhou, X. Topological mode switching in modulated structures with dynamic encircling of an exceptional point. *Proc. R. Soc. Math. Phys. Eng. Sci.* **477**, 20200766 (2021).
175. Yu, F., Zhang, X.-L., Tian, Z.-N., Chen, Q.-D. & Sun, H.-B. General Rules Governing the Dynamical Encircling of an Arbitrary Number of Exceptional Points. *Phys. Rev. Lett.* **127**, 253901 (2021).
176. Denner, M. M. *et al.* Exceptional topological insulators. *Nat. Commun.* **12**, 5681 (2021).

**Acknowledgments.** K. D. and G. M. thanks Zhen Li, Wei Wang, and Mengying Hu for helping in preparing the figures, and Ruo-Yang Zhang for discussions. C. F. thanks Zhesen Yang and Kai Zhang for discussions. G. M. is supported by the National Natural Science Foundation of China (11922416), the Hong Kong Research Grants Council (RFS2223-2S01, 12302420, 12300419, 12301822). K. D. is supported by the National Natural Science Foundation of China (12174072) and Natural Science Foundation of Shanghai (21ZR1403700). C. F. is supported by the Ministry of Science and Technology of China (2016YFA0302400) and the Chinese Academy of Sciences (XDB33000000).

**Author contributions.** All authors surveyed the literature and prepared the manuscript.

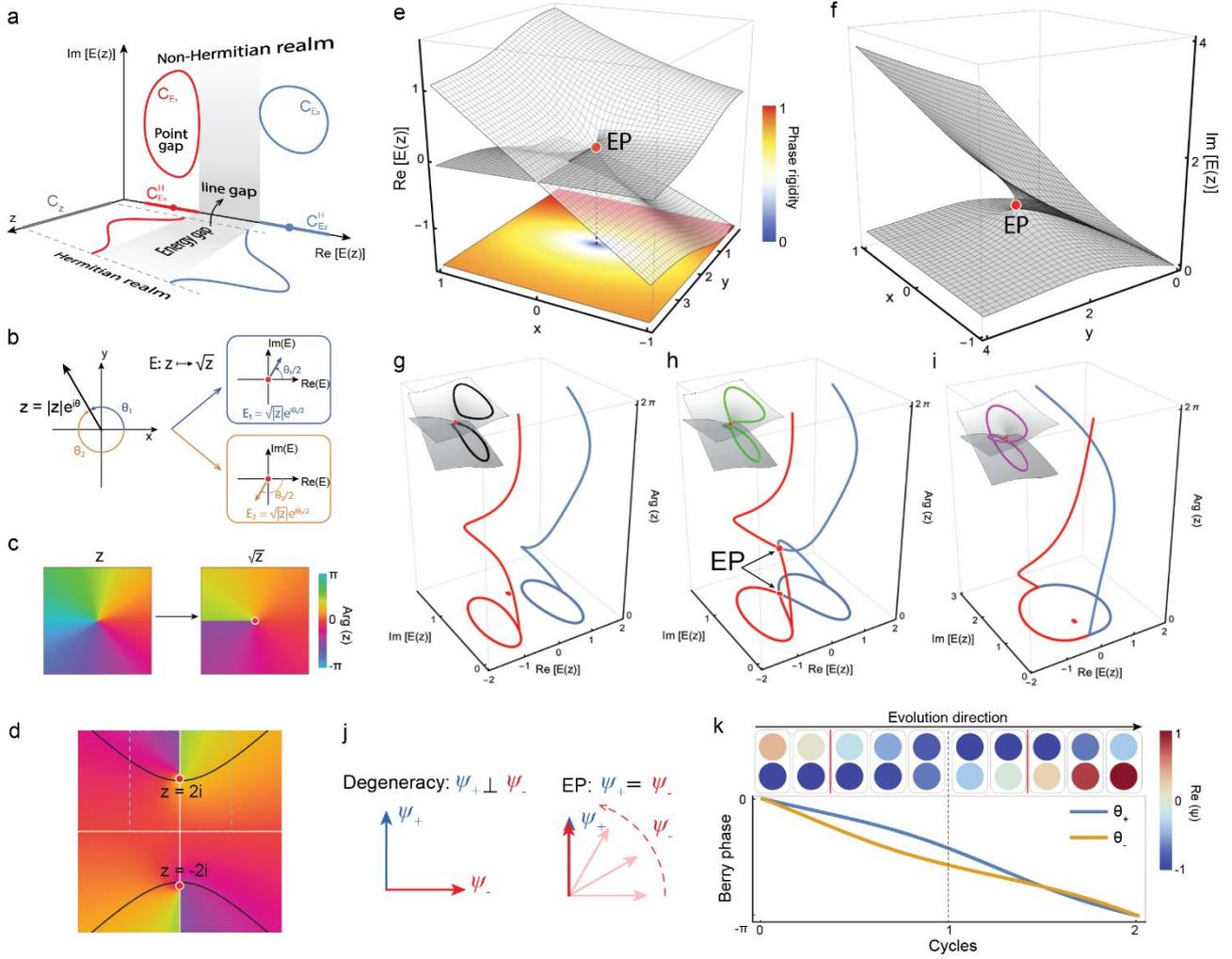

**Figure 1 | Eigenvalues and eigenvectors in non-Hermitian systems.** (a) Extension of real eigenvalues (Hermitian energy bands) to the complex energy plane. Here, a two-level system with $z$ as some system parameter(s) is shown. The Hermitian bands are entirely real and are separated by an energy gap. In the non-Hermitian realm, the two bands form loops on the complex plane. Two different cases are shown. The bands each forms closed loops with enclosing a finite spectral area, which is sometimes called a point gap. The Hermitian energy gap becomes a line gap in this case. Note that, in general, a line gap does not need to be perpendicular to the real-$E$ axis. (b) The square root for a complex number $z$ represented as a one-to-two map. (c) The arguments of $z$ and $\sqrt{z}$, where the branch cut represents a discontinuity in the argument map. (d) The argument of $E_{+,-}(z)$. The black and white lines represent $\text{Re}(\Delta)=0$ and $\text{Im}(\Delta)=0$, respectively, and their intersection points are the two exceptional points (EPs) marked by the red dots. (c, d) share the same colorbar. (e) Real parts and (f) imaginary parts of the eigenvalues near the order-2 EP at $z=+2i$. The model used here is described by Eq. (3), with $z = x + iy$. (g-i) The eigenvalue trajectories driven by $z = 0.8e^{i\theta} + z_0$ with $\theta$ from 0 to $2\pi$. (g) $z_0 = 0$ such that the loop does not enclose the EP. (h) $z_0 = 1.2i$ such that the loop

intersects with the EP. (i) $z_0 = 2i$ such that the loop encloses the EP. The insets of (g–i) show the eigenvalue evolutions as functions of $z$ on the real parts of $\mathcal{M}_E(z)$. (j) Eigenvectors are orthogonal at a two-fold Hermitian degenerate point. In contrast, non-Hermitian eigenvectors are identical at an EP. Phase rigidity shown in (e) is useful for quantifying the splitting of eigenvectors near an EP. The example here is described by Eq. (2). It vanishes at the EP and is unity at a distance far from the EP. (k) Parallel transport of $|\psi_+(z)\rangle$ around an EP at $z = 2i$. The upper row plots $|\psi_+\rangle$ at 10 chosen positions, where the colors indicate the real parts of the two entries. The two branch cuts are indicated by the red lines, across which the state changes parity. The state is recovered after two complete cycles, but the end state has a phase factor of $-\pi$ relative to the initial state, which is the Berry phase. The lower panel shows the accumulation of geometric phases for both $|\psi_+\rangle$ and $|\psi_-\rangle$ along the parallel transport. Note that only the difference between initial and terminal phases is gauge invariant, whereas the intermediate values are dependent on the winding path.

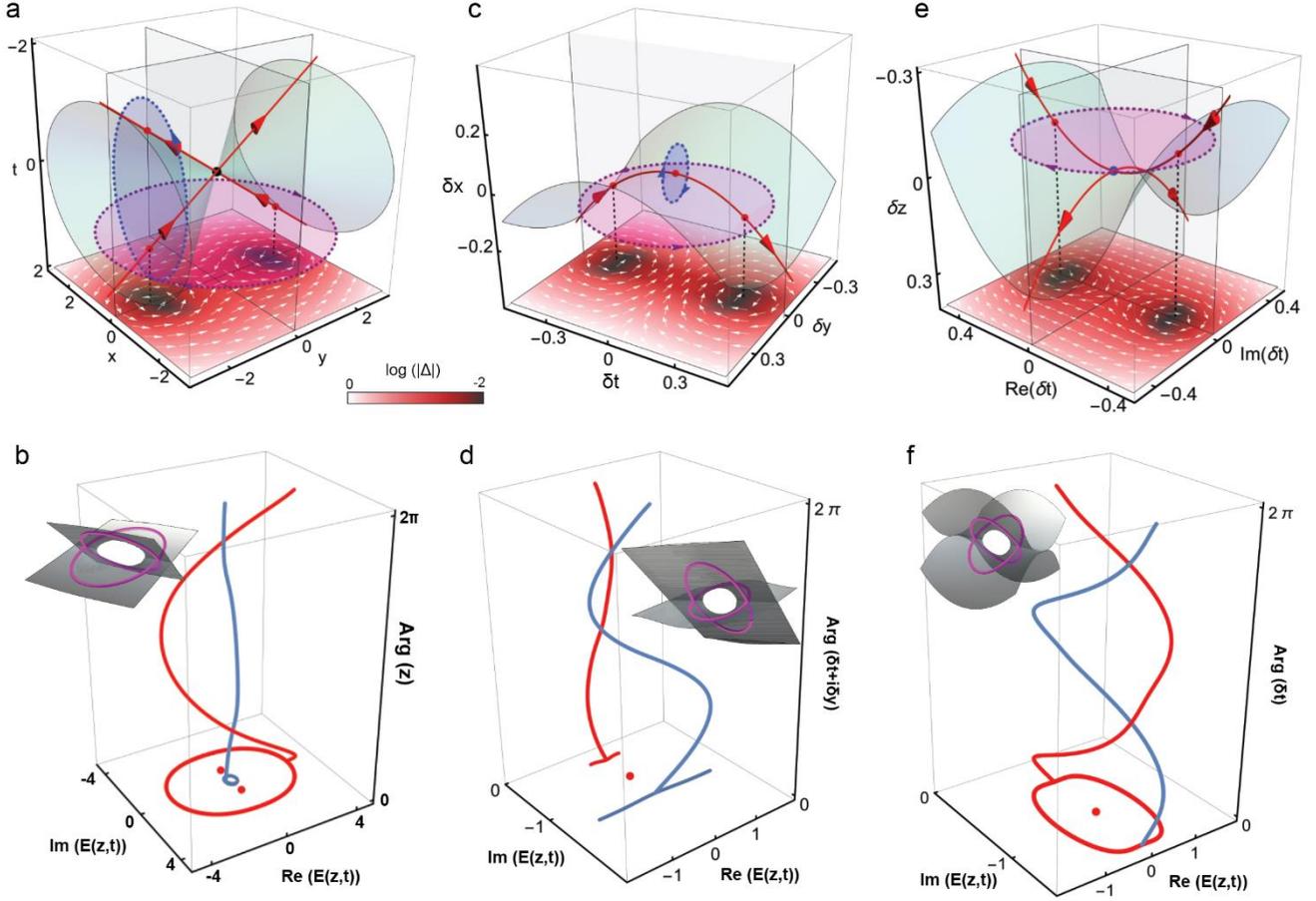

**Figure 2 | Exceptional lines and arcs**. (a) Two exceptional arcs (EAs), denoted by solid red lines, formed by regarding the hopping $t$ in the Hamiltonian (Eq. (3)) as a dimension. The red arrows indicate the orientations of the EAs, which are determined by the discriminant fields shown on the colormap in the bottom plane. The black dot at the crossing is a diabolic point. The dashed purple and dashed blue loops are nonhomotopic loops. (b) The eigenvalue braids along the dashed purple loop in (a). The inset shows the corresponding trajectories on the eigenvalue (real part) Riemann surface. (c) and (e) respectively show different scenarios of parabolic EA(s) in different parametric spaces. The blue dot in (e) where the two parabolas kiss is an EP with EWN of -2. The eigenvalue braids driven by a cycle of the purple loop in (c) and (e) are respectively depicted in (d) and (f). The azure and grey surfaces in (a), (c), and (e) are the surfaces when the real and imaginary parts of the discriminant vanish, and the EAs are their intersections. The colormaps in the bottom plane of (a), (c), and (e) show the norm of the discriminant in the log scale (base 10), and the arrows represent the local direction of the discriminant fields.

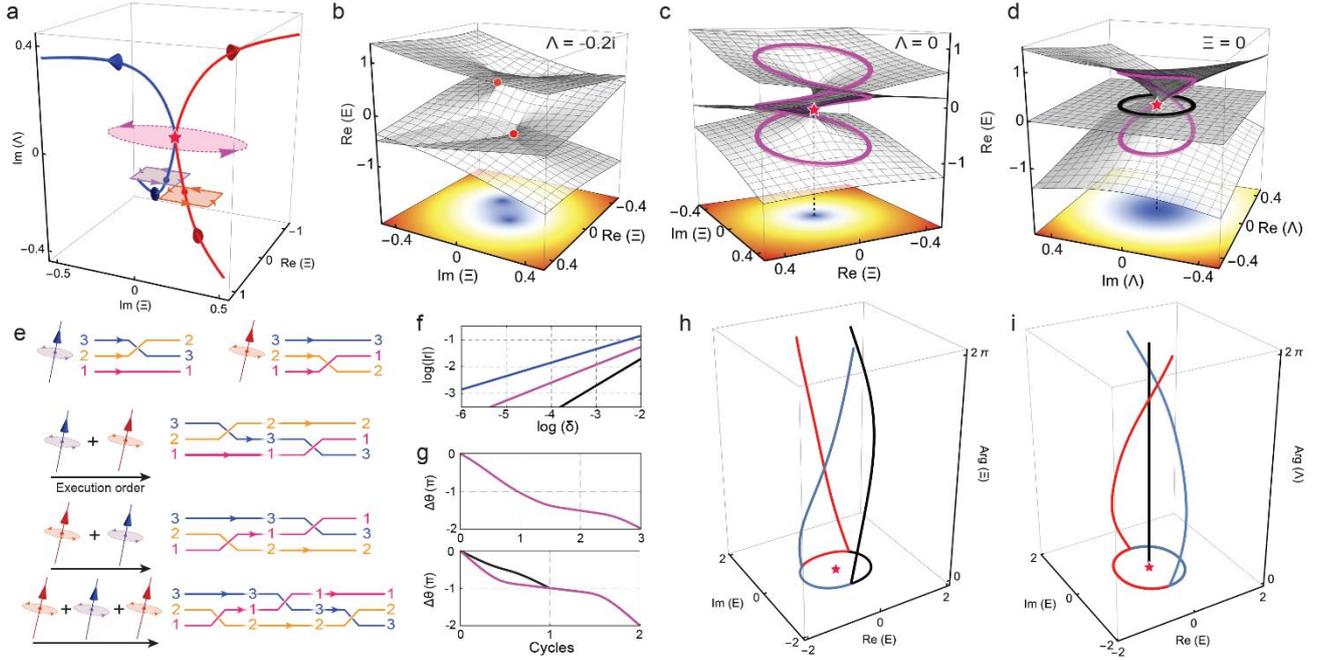

**Figure 3 | A higher-order exceptional point (EP) and its hybrid topological characteristics.** (a) A three-level non-Hermitian system exhibiting two order-2 exceptional arcs (EAs) (solid blue and solid red curves), which are formed by different pairs of bands. They converge to form an order-3 EP (the red star). (b, c) The real parts of the eigenvalue Riemann surface in the complex-$\Xi$ plane with $\Lambda = -0.2i$ (b) and $\Lambda = 0$ (c). (d) The eigenvalue Riemann surface in the complex $\Lambda$ plane with $\Xi = 0$. In (b-d), the color maps on the bottom plane show the phase rigidity of the middle state. (e) Non-Abelian state permutations enabled by the spectral topology. Encircling the blue (red) EA once exchanges states 2 and 3 (1 and 2). Concatenating the loops in different orders yields different end states. The eigenvalue braids in (h) and (i) are driven by a cycle of the loop shown in (c) and (d), respectively. The corresponding Berry phases are plotted in (g). (f) Double-log (base 10) plots of phase rigidities as function of the detuning parameters (denoted $\delta$). The magenta and black lines respectively correspond to the order-3 EP in (c) and (d), and the blue line corresponds to an order-2 EP in (b).

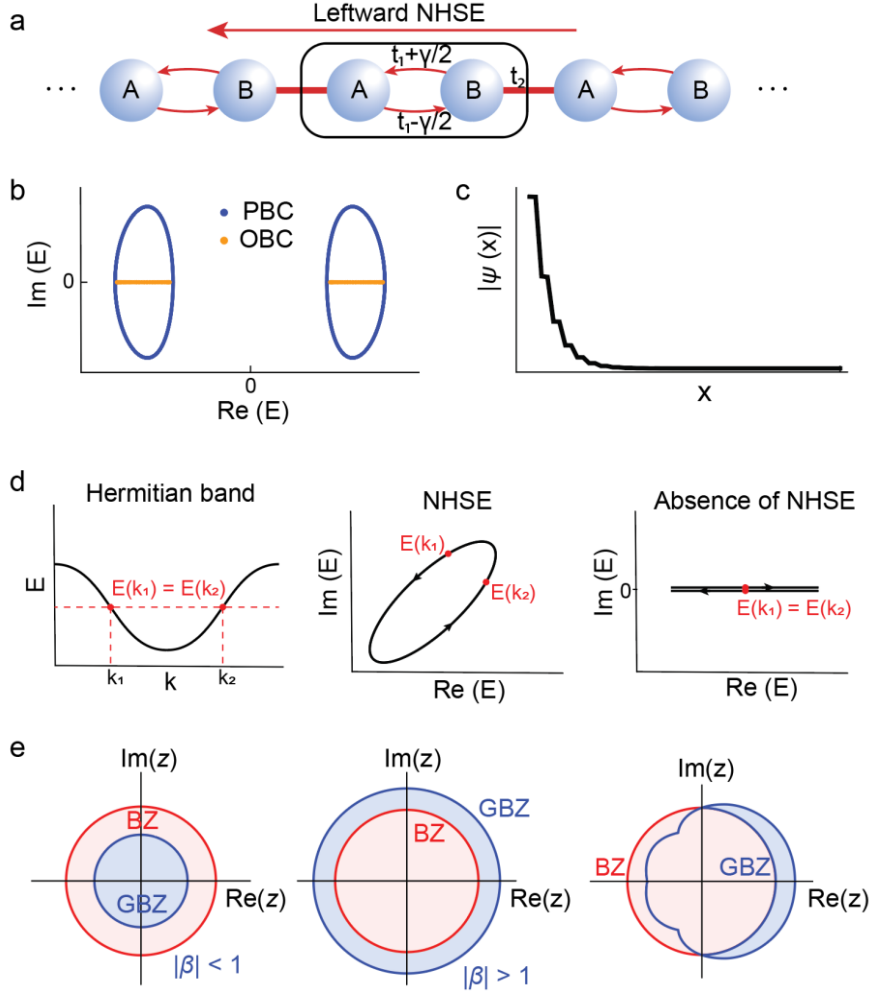

**Figure 4 | The non-Hermitian skin effect (NHSE) in a 1D lattice.** (a) The non-Hermitian SSH model with nonreciprocal intracell hopping, where the black box indicates a unit cell. (b) The complex spectra under a periodic boundary condition (PBC) (blue) and an open boundary condition (OBC) (orange). Here, the system is line-gapped and topologically trivial. Under the PBC, the two bands are separated by a line gap and each band forms a point gap. Under the OBC, the point gap is trivialized, and the spectra shrink to two lines on the real axis. (c) The norm of a skin mode. (d) In a Hermitian band (left), it is generically possible to find two (or more) different $k$ points with an identical energy, thereby allowing the formation of standing-wave solutions in an open-boundary lattice. In a non-Hermitian band forming point gaps enclosing a spectral area (middle), each $k$ maps to a unique complex energy. Hence OBC solutions cannot be constructed from PBC solutions via superposition, and the NHSE is the consequence. However, in the special case that a non-Hermitian band forms an arc without an interior (right), the NHSE is absent. (e) The Brillouin zones (BZs) and general BZs on the complex-$z$ plane. Here, $z = e^{ik}$ for the BZs and $z = \beta = |\beta|e^{i\theta}$ for the general BZs.

**Box 1. Non-adiabatic transitions in non-Hermitian dynamic evolutions**

Dynamic evolution is regarded as adiabatic in Hermitian systems with a discrete spectrum and a sufficiently slow variation in the value of their parameter(s), as the excited state remains at the same eigenvalue surface, and thus the parallel transport of eigenvectors is achieved. However, in some cases, such evolution can undergo a non-adiabatic transition[160] that converts a system into a "preferred" state; that is, a state with a higher gain or a lower loss. Such non-adiabatic transitions are asymmetric and dependent on both the initial state and the starting point (marked by the red stars in the panel) of the evolution, and are fundamentally related to the Stokes phenomena of asymptotics[161] and stability effects[162]. These non-adiabatic transitions occur because non-Hermiticity, such as gain or loss, tends to magnify any non-adiabatic imperfections during dynamic evolution. For example, in a system with a gain, a slight increase in the gain state will accumulate and amplify over time, such that the system is ultimately dominated by the gain. Therefore, a non-adiabatic transition invariably converts a system to its preferred state. It should be noted that such transitions are underlain by the geometry of the eigenvalue manifold, and may occur even when the evolution path does not enclose an exceptional point[163,164]. However, if the preferred state is chosen as the initial state, the non-adiabatic transition does not occur. Such effects have been explored for chiral mode switching-based wave manipulations[164–175].

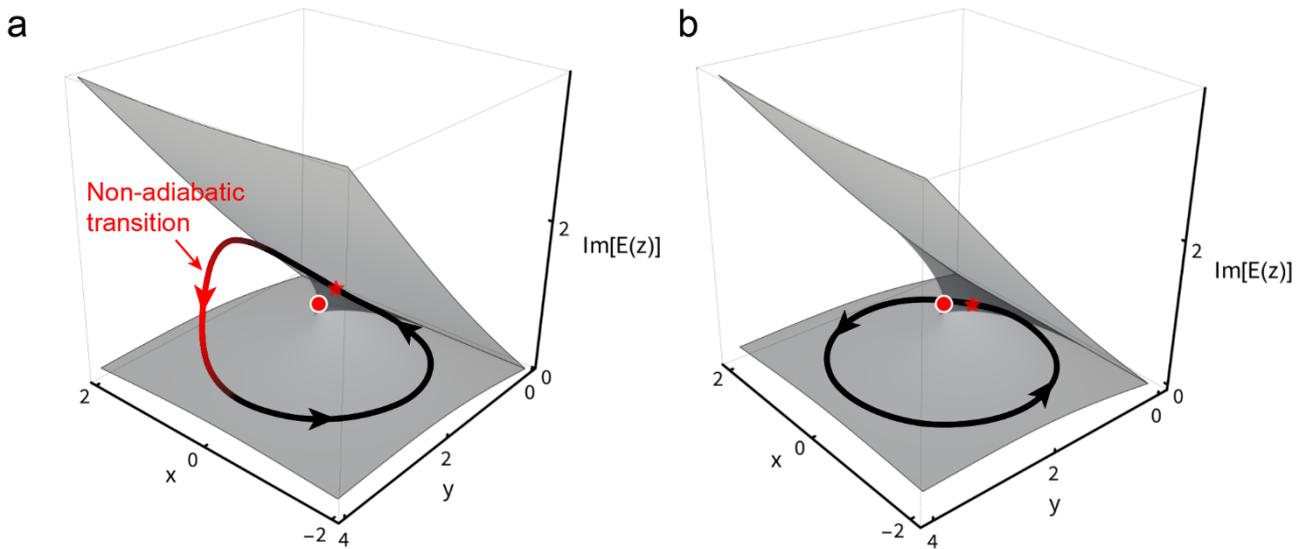

**Box 2. Point-gap topology of non-Hermitian Bloch bands**

Here, we discuss the roles of point gaps in the defining topological classifications of non-Hermitian periodic systems. For a given band system $\{E_i(\boldsymbol{k}), \boldsymbol{k} \in \text{BZ}\}$, consider the regions not occupied by the bands, that is $\bar{E} = \mathbb{C}\backslash[\cup_{i=1,\ldots,N} E_i, E_i(\boldsymbol{k} \in \text{BZ})]$. The bands break $\bar{E}$ into $M+1$ regions that are disconnected from each other, labeled as $\bar{E} = \bar{E}_0 \cup \bar{E}_1 \cup \cdots \cup \bar{E}_M$, where $M = \pi_0(\bar{E}) - 1$. Here we define $\bar{E}_0$ as the region containing $\infty$, and $\bar{E}_i$ for each $i \neq 0$ is called a point gap. If $H$ contains any symmetry, only those $\bar{E}_i$'s that are invariant under all symmetries can be called point gaps. This definition is equivalent to the one introduced in "Complex eigenvalues and spectral topology," but here we focus on the disconnected spectral areas instead of the eigenvalue trajectory $C_E$. Also, because $\boldsymbol{k}$ takes value in the entire BZ, in dimension larger than 1, it is possible for a band to fill an entire spectral area, as shown in the Figure below.

For a given point gap $\bar{E}_i$, choose one point $E_r \in \bar{E}_i$ that is invariant under all symmetries. Then consider the new Hamiltonian $\tilde{H} = H - E_r$, which belongs to the same symmetry class, and is invertible for all $\boldsymbol{k} \in \text{BZ}$. The latter is because the definition of point gap ensures that $E_0$ does not belong to any band, so that $\det(H - E_r) \neq 0$. For any invertible square matrix, a unique polar decomposition exists so that $\tilde{H}(\boldsymbol{k}) = U(\boldsymbol{k})P(\boldsymbol{k})$, where $U(\boldsymbol{k})$ is unitary and $P(\boldsymbol{k})$ is positive-definite and Hermitian. The classification of $H(\boldsymbol{k})$ is then converted to the classification of $U(\boldsymbol{k})$, which is a solved problem in the tenfold way. For example, consider a 1D one-band Hamiltonian $E(k)$, which forms a loop in the complex plane, dividing the plane into the region outside $\bar{E}_0$ and the one inside $\bar{E}_1$. Choose any point $E_0 \in \bar{E}_{0,1}$ we have $\tilde{H}(k) = E(k) - E_r$. The polar decomposition goes $U(k) = \frac{E(k)-E_r}{|E(k)-E_r|}$ and $P(k) = |E(k) - E_r|$. The classification of $U(k)$ is just the winding number, so we recover the EWN formula (Eq. (2)). Obviously, if $E_r \in \bar{E}_0$, then $\mathcal{W}_E = 0$ and for $E_r \in \bar{E}_1$, $\mathcal{W}_E \neq 0$. From this example, we see that the point gap topology depends on the choice of the reference point $E_r$, but different choices of $E_r$ do not change the point gap topology as long as $E_r$ belongs to the same $\bar{E}_i$. In other words, the point-gap topology is an extended version of spectral topology, which has been discussed above. According to refs. [120,121], in 1D, nontrivial spectral topology, i.e., the EWN, of the point gap $\bar{E}_i$ corresponds to the skin modes in the OBC spectrum. It is interesting to ask what point gap topology corresponds to in higher dimensions. While some special cases have been discussed[176], a general picture of bulk-edge correspondence has not been established.

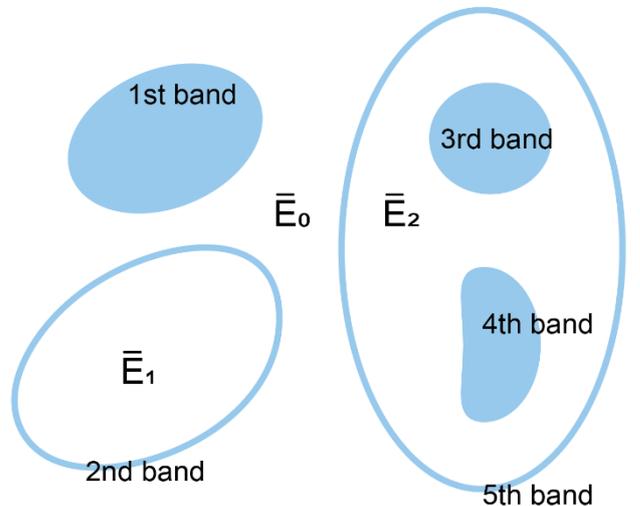